\documentclass{article}
\usepackage[tbtags]{amsmath}
\usepackage{amssymb}
\usepackage{amsthm}
\usepackage{array}
\usepackage{xy}

\usepackage[iso-8859-7]{inputenc}
\usepackage{graphicx}

\title{Cosmological vacuum selection and metastable susy breaking}
\author{Ioannis Dalianis and Zygmunt Lalak\footnote{e-mail addresses: Ioannis.Dalianis@fuw.edu.pl,\, Zygmunt.Lalak@fuw.edu.pl} \\
Institute of Theoretical Physics, Faculty of Physics, University of Warsaw\\
ul. Ho\.za 69, Warsaw, Poland}

\begin{document}

\date{}
\maketitle

\abstract{We study gauge mediation in a wide class of  O'Raifeartaigh type models  where supersymmetry breaking metastable vacuum is created by gravity and/or quantum corrections. We examine their thermal evolution in the early universe and the conditions under which the susy breaking vacuum can be selected. It is demonstrated that thermalization typically makes the metastable supersymmetry breaking cosmologically disfavoured but this is not always the case. Initial conditions with the spurion displaced from the symmetric thermal minimum and a small coupling to the messenger sector  can result in the realization of the susy breaking vacuum even if the reheating temperature is high. 
We show that this can be achieved  without jeopardizing the low energy phenomenology. 
In addition, we have found that deforming the models by a supersymmetric mass term for messengers in such a way that the susy breaking minimum and the susy preserving minima are all far away from the origin does not change the conclusions. 
The basic observations  are expected to hold also in the case of models with an anomalous $U(1)$ group.}

\section{Introduction} \normalsize

As the period of testing of low energy supersymmetry at the LHC comes near, the question about the true nature of supersymmetry breakdown and its transmission to the visible sector becomes increasingly important. As an alternative to the well studied case of purely gravitational/moduli  mediation, the gauge mediation of supersymmetry breaking 
\cite{Dine:1981za,Dimopoulos:1981au,Dine:1981gu,AlvarezGaume:1981wy,Dine:1982zb,Dimopoulos:1982gm,Nappi:1982hm, Dine:1993yw,Dine:1994vc,Dine:1995ag,Giudice:1998bp} with  the gravitino mass in the GeV range appears 
to be  a theoretically well supported possibility.  It naturally avoids excessive flavour-changing phenomena and predicts gravitino as the LSP, opening a perspective for distinct and promising phenomenology  \cite{Ibe:2007km, Heckman:2009bi} consistent with the standard cosmology \cite{Olechowski:2009bd}. 
However, to reliably study  gauge mediation one needs to take into account a complete theory, with the dynamics of the supersymmetry breaking sector coupled to messengers, and further to the visible sector. 
A simple and calculable supersymmetry breaking sector  can be described  by a variety of  
O'Raifeartaigh-type models, see, e.g.~\cite{O'Raifeartaigh:1975pr,Intriligator:2006dd,Kitano:2006wz,Murayama:2006yf,Cheung:2007es,Heckman:2008rb}. 
The renewal of the  interest in this class of models  is linked to the existence of metastable supersymmetry
breaking vacua in models with an approximate $R$-symmetry \cite{Intriligator:2007py}, 
although the idea of metastability in model building of supersymmetry breaking is not new, see \cite{Ellis:1982vi, Dimopoulos:1997ww}.
 There are two interesting aspects of such scenarios related to gravity. 
First of all, the gravitational mediation is always present and it is legitimate to ask to what extent one can mix the two channels of mediation. 
This question has been studied at length in~\cite{Lalak:2008bc}, and later in~\cite{Dudas:2008qf}. 
Secondly, one should ask about the cosmological history of such models. The existence of many competing vacua - supersymmetric and non-supersymmetric ones - poses the question of how natural it is for the complete theory to settle down into the phenomenologically relevant 
vacuum with broken supersymmetry. This has been discussed extensively in models based on the Intriligator, Seiberg, Shih (ISS) construction~\cite{Abel:2006cr,Craig:2006kx,Fischler:2006xh,Abel:2006my}. 
In this work we are going to discuss a wide class of models, different from the ISS,  where stabilization of the spurion field can be traced to the corrections to the 
K\"ahler potential. The models display a thermal evolution different from that of the ISS.
We shall deal in some detail with models where stabilization is due to a combined effect of K\"ahler potential and gravity corrections, and also with models 
where the stabilization occurs even in the global limit. These models have the property  that the metastable minimum lies further away from the origin than the supersymmetry preserving global minima, which is  opposite to the vacuum structure of the ISS-like models where the metastable minimum lies in the vicinity of the origin. The supersymmetry breaking vacuum of the theory here is metastable in the low energy limit and it resembles the type-II supersymmetry breaking vacua of \cite{Cheung:2007es}  which are phenomenologically viable (see also  \cite{Komargodski:2009jf}).  However, the question of the metastable vacuum selection during the evolution of the universe is  acute and challenging. We shall study thermal corrections in these models and examine effects due to  a possible coupling of the spurion to an inflaton. Eventually we shall  discuss  the question of how the metastable minimum can become more attractive.

\section{Outline} \normalsize
For convenience of the reader we present here  an outline of the forthcoming results.  
The models we study have a superpotential
\begin{equation}
W=\mu^2 S -\lambda S q \bar{q} \pm M q \bar{q} +c, 
\end{equation}
where $S$ is a gauge singlet responsible for supersymmetry breaking and $q$, $\bar{q}$ are messenger fields, a dimensionful constant $c$ has been included to account for cancellation of the cosmological constant. The 
general form of the K\"ahler potential reads
\begin{equation}
K=|S|^2- g_4|S|^4-g_6|S|^6.
\end{equation}
The $g_4$ and $g_6$ are dimensionful parameters 
of the form $g_4 = \epsilon_4 \Lambda ^{-2}$ and $g_6 = \epsilon_6 \Lambda^{-4}$ where $\epsilon_4$ and $\epsilon_6$ coefficients of unspecified sign. $\Lambda$ is the cut off scale of the theory which may be interpreted as the lowest mass scale of the particles which are integrated out. The coefficients can be rather different from each other, hence we write $|g_4|\equiv 1/\Lambda^2_1$ and $g_6 \equiv 1/\Lambda^4_2$. The first case we examine is  $M=0$, $g_6$ negligible and $g_4=1/\Lambda^2_1\equiv 1/\Lambda^2$. This model,  discussed  by Kitano \cite{Kitano:2006wz}, breaks supersymmetry in a metastable vacuum thanks to the presence of gravity which stabilizes this false-vacuum along the $S$-direction at a value of order $S \simeq \Lambda^2/M_P$. The supersymmetry preserving flat direction lies close to the origin at $S \simeq 0$ and $q\bar{q}= \mu^2/\lambda$. It is possible to stabilize the supersymmetry breaking vacuum even in the global supersymmetry limit for a K\"ahler potential with $g_4=-1/\Lambda^2_1$ and $g_6=1/\Lambda^4_2$ and the scale $\Lambda_2$ smaller - but not much smaller - than $\Lambda_1$. This is the second case. Otherwise, when $\Lambda_2$ is significantly smaller   than $\Lambda_1$, we have the third case and the theory resembles the first one.
%where supersymmetry breaking vacuum disappears in the global limit. 
If one neglects the dimensionfull constant $c$, the above three cases exhibit a $U(1)_R$ symmetry. In the first and the third  case it is the presence of this constant, visible only to gravity, that shifts the metastable vacuum far from the %tachyonic messenger-mass 
origin.  

When $M\neq 0$ we have the interesting effect of swapping the position of the supersymmetry breaking and supersymmetry preserving vacua, with the supersymmetry minimum lying at $S= \mp M/\lambda$ and $q\bar{q}=\mu^2/\lambda$. Due to the bare messenger mass the theory now lacks $U(1)_R$ symmetry even in the global supersymmetry limit. 

Since in these theories the phenomenologically preferred vacuum is the false one, it is important  to examine the thermal evolution of the system. 
%Moving back at the time direction we know our universe becomes hot and enters a phase of thermal equilibrium. 
In the early, hot, universe thermal effects change radically the vacuum structure. Thermally induced mass terms become large and 
% compensate the tachyonic messenger masses and 
the effective potential develops a minimum at $q=\bar{q}=0$ and $S \simeq \mp M/\lambda$.  At a temperature $T_{cr} \simeq 2\mu/\sqrt{\lambda}$  one finds a second order phase transition towards the supersymmetry preserving vacua for all the above cases. Hence, one expects that if the universe experienced reheating temperatures $T_\text {rh}>T_{q}=m_q/20$, i.e. messengers were thermalized, these theories are cosmologically disfavoured. In the case $T_\text {rh} < T_{q}$ the supersymmetry breaking vacua can be realized only if the system of fields is initially displaced from the finite temperature effective minimum $S= \mp M/\lambda$ (which is the origin in field space for $M=0$).  

However, taking into account the  inflationary phase, it is legitimate to assume  initial values for the scalar fields away  from the origin.  Displaced initial value for the spurion $S$ is the first necessary condition to avoid a fall into a supersymmetric minimum. The second condition is that the reheating temperature has to be bounded. It is demonstrated that the condition $T_\text{rh}<T_q$ can be circumvented. The supersymmetry breaking vacua appear at a temperature squared $T^2_S \sim c\mu/(M^2_P\sqrt{\lambda^3})$ with a small quantitative difference in the second case. This is lower that $T_{cr}$ but for a weak coupling $\lambda$ it can be larger than the $m_q/20$. Moreover, large reheating temperature $T_\text{rh}>T_S$ becomes acceptable in theories where the supersymmetry breaking and messenger sectors are very weakly coupled: $\lambda \ll 1$. The reason is that the induced thermal mass $\lambda T$ of the spurion field $S$ is too  weak to drive the $S$-field to the origin no matter how large the reheating temperature is. The Hubble friction freezes the field to its post-inflationary value until  the temperature drops down to the value of $T_S$  when the susy breaking vacuum appears. These constitute an improvement in model building allowing these theories to be accommodated in standard cosmological scenarios without resorting to scenarios of low reheating temperature and non-thermal evolution of the spurion and messenger fields.

In order to select the supersymmetry breaking vacuum  the system of fields needs to be aligned in the complex $S$-plane i.e. $q=\bar{q}=0$. This is expected to happen thanks to the thermal mass the messengers receive from MSSM degrees of freedom. Also, the fact that large vevs for the messengers make the SM gauge bosons heavy can result in a trapping effect on the fields $q$, $\bar{q}$ in the origin. On the other hand the spurion must be always outside the tachyonic origin of the complex $S$-plane.

Small couplings don't destabilize the supersymmetry breaking vacuum. The metastable vacuum is relegated further away from the supersymmetry preserving vacua making the - anyway small - tunneling probability completely negligible. Large values of the metastable vacuum, e.g. $S \simeq \Lambda^2/M_P$ in the first case, increase the contribution of gravity to  the soft masses. Asking for gauge domination results in an upper bound on the cut-off scale $\Lambda$. We conclude that a theory with very weakly coupled together supersymmetry breaking and  messenger sectors can be a cosmologically favourable theory which promotes a GeV range gravitino. It allows high reheating temperatures thus making it easier to accommodate leptogenesis scenarios. Of course, the later requires that the spurion does not cause a late entropy production.

We concentrate our study on the dynamics of the system composed of the spurion field $S$ plus the messengers $q$, $\bar{q}$. Generic effects of the MSSM degrees of freedom are taken into account when we discuss  thermalization of various  fields. We consider also the inflaton sector as a possible origin  of the initial vevs of $S$, $q$ and $\bar{q}$. It is interesting to note, that near the supersymmetric minimum the messengers acquire vevs which result in large masses of the SM gauge bosons. These bosons stay massless at the origin and at the supersymmetry breaking minimum, where they contribute to the thermal corrections. In this paper we assume that the additional contributions do not change the critical temperature $T_{cr}$ significantly, and consider their  possible role in the process of thermalization in Sections 6.2 and 7.  
%The MSSM and the UV completion of the model we study could modify some of the numerical results. 
%but generally we don't expect significant changes on the thermal history. 
%This is an interesting possibility which will be presented in a subsequent letter.
%but We note that Standard Model degrees of freedom have not directly taken into account. The SM degrees of freedom contribute to the thermal mass of the messengers. This effect can be important, especially at the limit of small coupling $\lambda$ between the spurion and the messengers. Large thermal mass for messengers compensate the tachyonic tree level masses until lower temperatures than one would expect simply from the coupling $\lambda S q \bar{q} $. This fact can decrease the $T_{cr}$ but not change significantly the thermal history. Actually, we consider the effects from MSSM fields in the section 6.2. There, it is shown that the spurion field generally cannot thermalize with MSSM degrees of freedom and hence, it cannot induce a thermal mass on the messengers.The thermal history of an extended susy breaking sector to the UV and IR region will be presented in subsequent work. Even though some results quantitatively may differ the main conclusions of this work hold.
\\
\\
The paper is organized as follows. After a brief review, in Section 3, of some general properties of gauge mediation models, we present in Section 4 a simple but elegant model of gravitational gauge mediation. We discuss in detail its thermal evolution assuming initially that the hidden/messenger sector can achieve thermal equilibrium. In the next step, in Section 5, we generalize the allowed corrections to the K\"ahler potential in such a way, that a stabilization and supersymmetry breaking can be achieved without gravity or via the balance between gravitational effects and the order 6 correction to the K\"ahler potential of the spurion. In the case that the stabilization survives in the absence of gravity, neglecting messengers, a spontaneous $R$-symmetry and supersymmetry breaking is observed in the global limit via a second order phase transition. We also describe the thermal evolution of these more general models. 
The conclusion is that, in case of messenger sector thermalization and localization of the spurion in the symmetric origin, the (closed) system of fields typically is dragged towards the supersymmetry preserving vacuum.

Next in Section 6, we discuss asymmetric, displaced  initial conditions for the cosmological evolution of the hidden sector. We consider inflationary fluctuations and a direct coupling of the spurion to the inflaton as a possible source of such initial conditions and comment on the evolution of the system. The conditions for thermalizations are also discussed. We find in Section 7 that when the messengers are weakly enough coupled to the spurion the supersymmetry breaking vacuum can be realized. Moreover, when the coupling takes  smallest allowed values, the reheating temprerature can be unbounded from above and, still, the attractive minimum the  metastable one. In Section 8, we assemble the constraints on the parameters of the models examined and conclude that a GeV range gravitino mass is cosmologically favoured.
Later, in Section 9,  we extend our study to a model that exhibits supersymmetry breaking vacuum lying closer to the origin than the supersymmetry preserving one. 
This is a simple extension of the earlier models: one simply  adds an explicit mass term for  messengers. After taking into account gravitational effects we conclude, that thermal evolution of this model is very close to that of the previous ones. Then, in Section 10, we consider briefly the widely studied ISS model and comment on the similarities and differences between this  model and the models discused earlier in the paper. Finally, we comment, in Section 11, on the setups where the spurion is charged under an additional $U(1)$ gauge symmetry and argue that the general analysis presented here applies also to this gauge invariant case. We summarize the results of this work in Section 12. Also, some general properties of the finite temperature effective potential are presented in the appendix.

\section{Gauge mediated susy breaking} \normalsize

Gauge mediation of supersymmetry breakdown from the hidden to the visible sector assumes that there exists  a nongravitational, although indirect, coupling between the two sectors. In this case there are messengers with the gauge interactions of the Standard Model which couple linearly in the superpotential  to the superfield $S$, the spurion,  whose F-term controls  supersymmetry breakdown:
\begin{equation}
W=\lambda S \bar{q} q \, ,
\end{equation}
where $\bar{q}$, $q$ are messenger superfields, which in the simplest case form respectively  $\bar{\bold{5}}$ and $\bold{5}$ of $SU(5)$, to preserve gauge coupling unification. At the scale $\langle S \rangle$ the Standard Model gaugino masses are given approximately by 
\begin{equation}
M_j=k_j\frac{\alpha_j}{4\pi}\frac{\langle F_S \rangle }{\langle S \rangle }, \ \ \ j=1,2,3
\end{equation} 
where $k_1=5/3$, $k_2=k_3=1$ and $\alpha_i$ are the three standard model gauge couplings. The scalar masses are given by 
\begin{equation}
\tilde{m}^2=2\sum^{3}_{j=1}C_jk_j \left ( \frac{\alpha_j}{4\pi}\right )^2\left(\frac{\langle F_S \rangle }{\langle S \rangle }\right)^2,
\end{equation} 
where $C_3=4/3$ for colour triplets, $C_2=3/4$ for weak doublets (and equal to zero otherwise) and $C_1=Y^2$ with $Y=Q_{el} -T_3$. To have squarks and gaugino masses of order $100$ GeV - $1$ TeV, we need $\langle F_S \rangle / \langle S \rangle = 10^2 - 10^3 \,$TeV. It is natural to assume that $\langle S \rangle $ is not larger than the unification scale 
of the order of $10^{16}$ GeV, which in turn implies $\sqrt{\langle F_S \rangle } < 10^{11}$ GeV. The ratio of gravitational\footnote{Throughout this paper we use $M_{P} =  2.4\times 10^{18}$GeV.} to gauge mediated contributions to soft masses is $\frac{\delta m_{gauge}}{\delta m_{grav}}  \simeq \frac{\alpha}{4\pi} \frac{M_{P}}{\langle S \rangle }$, which means  that the smaller $\langle S \rangle $, the stronger the gauge mediation dominance is. When a bare messenger mass is present, gaugino masses are given by 
$M_j=k_j \frac{\alpha_j}{4 \pi} \frac{\lambda\langle F_S \rangle}{M_\text{mess}}$ and $\frac{\delta m_{gauge}}{\delta m_{grav}} \simeq \frac{\alpha}{4\pi} \frac{\lambda M_{P}}{M_\text{mess}}$.

Ways of  realizing a (local) non-supersymmetric minimum for finite (but smaller than the Planck scale)  $S$ have been discussed in the literature. Of interest to us are O'Raifeartaigh models.  Their small mass scales, needed for realistic phenomenology, may be generated dynamically, or retrofitted, or may be generated with the help of the D-brane instantons.

\section{Gravitationally stabilized metastable SUSY breaking} \normalsize

A simple model proposed by Kitano \cite{Kitano:2006wz} achieves metastable susy breaking due to gravitational effects. The K\"ahler potential and the superpotential are
\begin{equation} \label{KitanoK}
K=S^{\dagger}S-\frac{(S^\dagger S)^2}{{\Lambda}^2}+q^\dagger q+\bar{q}^\dagger \bar{q}
\end{equation} 
\begin{equation} \label{KitanoW}
W=\mu^2 S-\lambda Sq\bar{q}+c.
\end{equation} 
The chiral superfield $S$ is a gauge singlet, while $q$ and $\bar{q}$ are the messenger fields which carry standard model quantum numbers and $\lambda$ is a coupling constant. The constant term $c$ does not have any effect if we neglect gravity interactions, but it is important for the cancellation of the cosmological constant. If we neglect the constant  $c$, the Lagrangian has an $R$-symmetry with charge assignments R($S$)=2, R($q$)=R($\bar{q}$)=0.

It is also necessary to estimate the perturbative quantum corrections to the K\"ahler potential coming from the interaction term $\lambda S q \bar{q}$ which may be more important than the gravity effect. At one loop level the correction is \cite{Kitano:2006wz, Intriligator:2006dd}
\begin{equation} \label{KitanoK-loop}
K_\text{1-loop}=-\frac{\lambda^2 N_q}{(4\pi)^2}S^\dagger S \log \frac{S^\dagger S}{Q ^2}
\end{equation} 
where $N_q$ the number of components in $q$ and $\bar{q}$ and $Q$ is the UV cut off scale. For example, $N_q=5$ if $q$ and $\bar{q}$ transform as 
$\bold{5}$ and $\bold{\bar{5}}$ under $SU(5)_{GUT}$. At low energies messengeres are integrated out and their effects are incorporated at (\ref{KitanoK-loop}). The cut off scale $\Lambda$ also originates from microscopic scale physics. For example, it may account for massive fields of the UV completion of the susy breaking sector which have been integrated out at the energy scales $E<\Lambda$ that we are considering here.

The scalar potential of the supergravity Lagrangian is given by 
\begin{equation} \label{}
V=e^G\left(G_S G_{S^\dagger}G^{S S^\dagger}+G_q G_{q^\dagger} G^{q q^\dagger}+G_{\bar{q}} G_{\bar{q}^\dagger} G^{\bar{q} \bar{q}^\dagger}-3\right)+\frac{1}{2}D^2,
\end{equation} 
where $G = K+\log(|W|^2/M^6_P)$ and $D^2/2$ represents the $D$-term term contributions that we neglect. 
The supersymmetric minimum is \footnote{We note that in the global susy limit we have the susy preserving \itshape flat direction \normalfont : $q\bar{q}=\mu^2/\lambda, \, S=0$.}
\begin{equation} 
q=\bar{q}=\sqrt{\frac{\mu^2}{\lambda}}-{\cal O}\left( \frac{c^{}}{\lambda^{}M^{2}_P}\right), \   \  \  S={\cal O}\left(\frac{c}{\lambda M^2_P}\right)
\end{equation}

Along the $S$ direction the potential simplifies to 
\begin{equation} \label{V-zero}
V(S)\simeq \mu^4-3\frac{c^2}{M^2_P}-2\frac{c}{M^2_P}\mu^2 (S+S^\dagger)+4\mu^4 \frac{|S|^2}{{\Lambda}^2} + \frac{\lambda^2 N_q}{(4\pi)^2} \, \mu^4 \log \frac{S^\dagger S}{Q^2},
\end{equation} 
which for $\lambda^2 N_q/(4\pi)^2 <(\Lambda/M_P)^2$ gives the non-supersymmetric minimum
\begin{equation}
  \left\langle S \right\rangle=\frac{c {\Lambda}^2}{2\mu^2M^2_P}.
\end{equation}
The parameters $c$ and $\mu$ are connected via the cancellation of the cosmological constant
$\mu^4 \simeq 3c^2/M^2_P\, .$
With the help of this condition the minimum can be written as
\begin{equation} \label{S-min}
\left\langle S \right\rangle\simeq \frac{\sqrt{3}\Lambda^2}{6M_P} \, .
\end{equation}
Supersymmetry is broken with $F_S = \mu^2$. One can see that in the global susy limit  $M_{P}\rightarrow \infty$ the minimum moves to $S\rightarrow 0$ and the metastable vacuum disappears. It is the presence of gravity that reveals the non-supersymmetric vacuum. 
The dominant terms in the potential, the tree level plus the one loop correction (\ref{KitanoK-loop}), up to 4th order in fields reads:
\begin{eqnarray}  \label{Kitano-tree} \nonumber
&V \simeq \mu^4-3\frac{c^2}{M^2_P}-2\frac{c}{M^2_P}\mu^2 (S+S^\dagger)+4\mu^4 \frac{|S|^2}{{\Lambda}^2}
%+(\mu^4-3\frac{c^2}{M^2_P})\frac{1}{M^_P}(|q|^2+|\bar{q}|^2)
- \lambda\mu^2(q\bar{q}+q^\dagger\bar{q}^\dagger)+2 \frac{c}{M^2_P}\mu^2 \frac{|S|^2}{\Lambda^2} (S+S^\dagger)& \\ \nonumber 
&%-2 \frac{c}{M^4_P} \mu^2  (S +S^\dagger)(|q|^2+|\bar{q}|^2)
+\lambda^2 |S|^2 (|q|^2+|\bar{q}|^2)+ \lambda^2 |q|^2 |\bar{q}|^2 + 
\frac{\lambda^2 N_q}{(4\pi)^2} \, \mu^4 \log \frac{S^\dagger S}{Q^2} & 
%&-4\lambda \mu^2\frac{|S|^2}{M^2_P\Lambda^2}(q\bar{q}+q^\dagger\bar{q}^\dagger)+4\mu^4 \frac{|S|^4}{{M^2_P\Lambda}^2}+ &
\end{eqnarray}
where $\mu^4\approx 3c^2/M^2_P$.

The mass matrices for  $S$ and $q$ are:
\begin{equation}
m^2_S \simeq \left(\begin{array}{cl}      
4\frac{\mu^4}{\Lambda^2} &  -\frac{\lambda^2N_q}{(4\pi)^2}\frac{\mu^4}{S^{\dagger \,2}}  \\     
-\frac{\lambda^2N_q}{(4\pi)^2}\frac{\mu^4}{S^2} & 4\frac{\mu^4}{\Lambda^2}   \\     
\end{array}\right),
%\end{math}
\,\,\,\,\,\, m^2_q \simeq
%\begin{math}
\left(\begin{array}{cl}      
\lambda^2 |S|^2 & -\lambda \mu^2  \\     
-\lambda \mu^2 & \lambda^2 |S|^2   \\     
\end{array}\right)
\end{equation}
and should be evaluated at the susy braking vacuum: $\left\langle  S \right\rangle \sim \Lambda^2/M_P$ and $q=\bar{q}=0$.
The susy breaking minimum is stable in the $q$, $\bar{q}$ directions when the determinant of the $q$-$\bar{q}$ mass matrix is positive which yields
\begin{equation} \label{vac-stab}
\lambda^2 \left\langle S \right\rangle^2 >  \lambda F_S \; \; {\rm i.e.} \;\; \Lambda^4/M^2_P> \mu^2/\lambda. 
\end{equation}
Thus, the susy breaking metastable minimum is further away from the origin than the susy preserving one. It is stable in the $S$-direction when $\lambda^2 N_q/(4\pi)^2 <{\Lambda}^2/M^2_P$ or roughly 
\begin{equation} \label{vac-stab-qua}
\lambda < \Lambda/M_P .
\end{equation}
This condition renders the perturbative quantum corrections harmless for the vacuum meta-stability. It says that the closer to the origin is the metastable vacuum, the smaller has to be (the square of) the coupling $\lambda$. 

Moreover, there is a phenomenological requirement that the gaugino masses should be of the order of $m_{\text{gaugino}}={\cal O}(100\ \text{GeV}-1\ \text{TeV})$. This fixes the relation between the parameters $\mu^2$ and $\Lambda$ as follows:
\begin{equation} \label{muLa}
\mu^2 = \left(\frac{\alpha}{4\pi} \right)^{-1} m_{\text{gaugino}} \left\langle S\right\rangle \simeq 10^{-14} \Lambda^2.
\end{equation}
With fixed gaugino masses we have two parameters: $\Lambda$ and $\lambda$. 
The two conditions (\ref{vac-stab}) and (\ref{vac-stab-qua}), necessary for  gravitational stabilization, can both be fulfilled for 
\begin{equation} \label{Lambda}
\Lambda>10^{-14/3}M_P .
\end{equation}
This lower bound on $\Lambda$ is high  enough to keep the metastable vacuum far away from the supersymmetric one and  to suppress sufficiently  the tunneling rate. The shape of the zero temperature potential around the minima is depicted in the Figure 1, the constraints on the parameters from the susy breaking vacuum stability are presented in the Table 1 and illustrated in the Figure 2.

\subsection{Gravitational gauge mediation at finite temperature} \normalsize

In the Kitano model the messengers $q$ and $\bar{q}$ carry standard model quantum numbers and we expect them to be in thermal (and chemical) equilibrium in the early universe. The chiral superfield $S$ of the secluded sector is coupled with a coupling $\lambda$ to the messenger sector and, in principle,  it can also achieve thermal equilibrium. Let's assume\footnote{In subsequent sections we shall examine whether this assumption holds.} the whole system to be in thermal equilibrium at temperature $T$. The interactions induce  a thermally corrected potential with a shape  different from the zero temperature one. According to the equation ($\ref{V-T}$) of the appendix, to obtain  the ${\cal O} (T^2)$ correction we should compute the relevant contributions to the traces of the squared mass matrices $TrM^2_s$ and $TrM^2_f$ (or the sum of their eigenvalues).
Also, for precise description of the evolution of the fields we need the ${\cal O}(T)$ contribution, although in the high temperature limit the ${\cal O}(T^2)$ dominates. The quadratic and the linear in $T$  part of the finite temperature corrected potential is
\begin{equation} \nonumber
V \supset \frac{T^2}{24}\left(tr\{M^2_{s}\}+tr\{M^2_{f}\}\right)-\frac{T}{12\pi} tr\{M^2_s\}^{3/2} ,   \, \,\,\,\,\,\,\,\,\,\,\,\,\,\,\,\,\,\ \; \; \; \; \; \; \; \;
\end{equation}
where, as we have verified, the non-canonical terms in the K\"ahler potential can be safely neglected. For $q=\bar{q}$ the mass matrix has two positive eigenvalues $\lambda_{1,2}=(M^2_S)_{1,2}$ (the negative eigenvalues give rise to the imaginary part of the potential which we neglect). 
%These two eigenvalues are approximately
%\begin{equation} \nonumber
%\lambda_1\approx 4\frac{\mu^4}{\Lambda^2}-2\frac{\mu^2c}{\Lambda^2 M^2_P}(S+S^\dagger)+2\lambda^2|q|^2
%\end{equation}
%\begin{equation} \nonumber
%\lambda_2\approx 2\lambda\mu^2+2\lambda^2|S|^2+2\lambda^2|q|^2.
%\end{equation}
The full thermally corrected potential reads:
\begin{equation} \nonumber
V \simeq \frac{c^2}{M^4_P}{\Lambda}^2-2\frac{c}{M^2_P}\mu^2 (S+S^\dagger)+4\mu^4 \frac{|S|^2}{\Lambda^2}- 2\lambda\mu^2|q|^2-4 \mu^2 \frac{c}{M^4_P} (S+S^\dagger) |q|^2 +2\lambda^2 |S|^2 |q|^2+ {\cal O}(S^3)+\lambda^2 |q|^4
\end{equation} 
\begin{equation} \nonumber
-\frac{\pi^2T^4}{90}N+\frac{T^2}{12}\left[4\frac{\mu^4}{{\Lambda}^2}+(S+S^\dagger)(4\mu^2\frac{c}{M^2_P}\frac{1}{{\Lambda}^2})+|S|^2(3\lambda^2)+|q|^2(6\lambda^2)\right]
\end{equation}
\begin{equation} \label{fullT-Kitano}
-\frac{T}{12\pi}\left[\left(4\frac{\mu^4}{\Lambda^2}-2\frac{\mu^2c}{\Lambda^2 M^2_P}(S+S^\dagger)+2\lambda^2|q|^2\right)^{3/2}+\left( 2\lambda\mu^2+2\lambda^2|S|^2+2\lambda^2|q|^2 \right)^{3/2}\right].
\end{equation}

\subsection{The minima and the evolution of the finite temperature scalar potential} \normalsize

In what follows we have in mind thermal expectation values of real fields unless stated otherwise. 
To start with we shall examine the shape of the finite temperature potential in various directions in the field space.
\\
\\ 
\bfseries \itshape The $q$-direction.  \normalfont
\\
Firstly, in the $q$-direction i.e. taking $S=0$, the (\ref{fullT-Kitano}) reads
\\
\begin{equation} 
V^q= \text{const}(T)-2\lambda\mu^2q^2+\lambda^2q^4+\frac{T^2}{2}\lambda^2q^2-\frac{T}{12\pi} \left[\left(4\frac{\mu^4}{\Lambda^2}+2\lambda^2q^2\right)^{3/2}+\left( 2\lambda\mu^2+2\lambda^2q^2 \right)^{3/2}\right].
\end{equation}
We can write it in a simpler form
\begin{equation} \label{q-dir}
V^q \simeq  \text{const}(T)+\frac{1}{2}m^2_q(T)q^2-{\cal O}(10^{-1})\lambda^3Tq^3+\lambda^2q^4.
\end{equation}
where we defined the effective $q$-mass as $m^2_q(T)\equiv\lambda^2T^2-4\lambda\mu^2$. The expectation value of the $q$ scalar field is obtained by minimizing the potential. For sufficiently high temperatures there is only one solution of $\partial V/\partial q=0$, namely 
\begin{equation}
q=0,
\end{equation}
and this is a minimum so long as $m^2_q(T)$ is positive. The effective mass changes sign from positive to negative (becomes tachyonic) at temperature $T_0$ 
\begin{equation} \label{T-0}
T_0=2\frac{\mu}{\sqrt{\lambda}}
\end{equation}
and we may write it as 
\begin{equation}
m^2_q(T)=-4\lambda\mu^2(1-T^2/T^2_0)
\end{equation}
We can see that the $\partial V/ \partial q=0$ of (\ref{q-dir}) has two more solutions: a maximum and a second (local) minimum when $m^2_q(T)\leq (9/16){\cal O} (10^{-2}) \lambda^4 T^2$ and this occurs when temperature drops below $T_1$ where
\begin{equation} \label{T-1}
T^2_1=\frac{T^2_0}{1-(9/16){\cal O}(10^{-2})\lambda^2}>T^2_0.
\end{equation}
The second minimum and the maximum are at
\begin{equation} \label{q-sol}
q_{\pm}(T)=\frac{\mu}{\sqrt{\lambda}} \left[\frac{{\cal O}(10^{-1})\lambda^{3/2}}{\mu}T\pm\left(1-\frac{T^2}{T^2_1}\right)^{1/2}\right].
\end{equation}
At $T=T_1$ the $q_{\pm}(T_1)$ is an inflection point. Below $T_1$ we have the formation of the second local minimum. At $T=T_{cr}$ this minimum becomes degenerate with the global minimum $q=0$. This happens when $m^{2}_q (T) \leq (1/2){\cal O}(10^{-2}) \lambda^4 T^2$. This relation gives a critical temperature slightly lower than $T_1$:
\begin{equation} \label{T-cri}
T^2_{cr}=\frac{T^2_0}{1-(1/2){\cal O}(10^{-2})\lambda^2}<T^2_1.
\end{equation}
We see that $T_0<T_{cr}<T_1$. Below the critical temperature the global minimum changes discontinuously from $q=0$ to $q_{+}(T_{cr})$. The origin $q=0$ becomes metastable until $T_0$ when the barrier that keeps the origin locally stable disappears and then $q=0$ becomes a local maximum. However, comparing the ($\ref{T-0}$), ($\ref{T-1}$) and ($\ref{T-cri}$) we see that they are only slightly different. Therefore, we can safely say 
\begin{equation}
T_1 \cong T_{cr} \cong T_0.
\end{equation}
In other words, the origin becomes tachyonic simultaneously with the appearence of a second asymmetric minimum, to a good approximation. This is equivalent to neglecting the term linear in temperature in equation ($\ref{q-dir}$). This phase transition is practically of second order \cite{Quiros:1999jp} and takes place at the critical temperature
\begin{equation} \label{T-cr}
T_{cr} \cong 2 \frac{\mu}{\sqrt{\lambda}}.
\end{equation}
The solution ($\ref{q-sol}$) gives the late-time minima in the $q$-direction: $q_{\pm}(T\rightarrow 0)=\pm \mu^2/\sqrt{\lambda}$, which are the supersymmetric minima of the tree level potential, while $q=0$ ends up as a local maximum.
\\
\\ 
\bfseries \itshape The $S$-direction.  \normalfont
\\
Secondly, in the $S$-direction, i.e. for $q=\bar{q}=0$ the potential (\ref{fullT-Kitano}) reads
\\
\begin{equation} \nonumber
V^S\simeq \text{const}(T)-4\frac{c}{M^2_P}\mu^2 S+4\mu^4 \frac{S^2}{{\Lambda}^2} +{\cal O}(S^3)
-\frac{\pi^2T^4}{90}N+ 
\end{equation}
\begin{equation}
+\frac{T^2}{12}\left[S(8\mu^2\frac{c}{M^2_P}\frac{1}{{\Lambda}^2})+S^2(3\lambda^2)
\right]-\frac{T}{12\pi}\sqrt{8}\lambda^3S^3.
\end{equation}
\\
The minimum along the $S$-direction is given by 
\begin{equation} \label{S-min0}
S_{min}(T)=\frac{4\frac{c}{M^2_P}\mu^2-\frac{2\mu^2c}{3\Lambda^2M^2_P}T^2}{8\frac{\mu^4}{\Lambda^2}+\frac{1}{2}\lambda^2T^2}.
\end{equation}
\\
We note the position of the minimum at different temperatures: \\
$\alpha$) for $T>{\Lambda} \  \Rightarrow \  S_{min}\simeq -\frac{4}{3}\frac{\mu^2c}{\lambda^2{\Lambda}^2M^2_P}$ , \\
$\beta$) for $T \sim T_{cr}=2\mu/\sqrt{\lambda} \  \Rightarrow \  S_{min}\sim \frac{c}{\lambda M^2_P}$ ,\\
$\gamma$) for $T \ll T_{cr}=2\mu/\sqrt{\lambda} \  \Rightarrow \  S_{min}\simeq \frac{4c\mu^2/M^2_P}{8\frac{\mu^4}{{\Lambda}^2}+\frac{1}{2}\lambda^2T^2}\rightarrow  \frac{\Lambda^2}{M_P} \ \text{as} \ T\rightarrow 0 $, \\
$\delta$) for $T=0  \Rightarrow \  S_{min}= \frac{\sqrt{3}}{6}\frac{\Lambda^2}{M_P}$. \\
\\
\bfseries \itshape The evolution of the $S-q$ system.  \normalfont
\\
Until now we have examined independently the evolution of the $S$ and $q$ directions. However, what we deal with  is a coupled $S$-$q$ system that could evolve in a different way. To see what actually happens we write the scalar potential truncated to the most relevant terms:
\begin{equation} \nonumber
V \simeq \text{const}-4\frac{c}{M^2_P}\mu^2S\left(1-\frac{T^2}{6\Lambda^2}\right)+S^2\left(\frac{4\mu^4}{\Lambda^2}+\frac{\lambda^2T^2}{4}\right)
+q^2\left(-2\lambda\mu^2+2\lambda^2S^2+\frac{T^2}{2}\lambda^2 \right)+\lambda^2q^4.
\end{equation}

The first remark is that the complete effective mass squared of the $q$ field is here $\partial^2 V(S,q,T)/\partial q^2\equiv m^2_{q,eff} \equiv \lambda^2 T^2-4 \lambda \mu^2+4\lambda^2 S^2$. For high enough temperatures  the condition $\partial V/\partial q=0$ is satisfied for $q=0$. On the other hand, due to the linear term, the high temperature minimum of the $S$ field is $S_{min}=-4\mu^2c/(3 \lambda^2 \Lambda^2 M^2_P)$. Thus, the effective mass of $q$ changes at high temperatures  and it has a non-zero contribution from the $S$ field. However, we see from  the zero temperature vacuum meta-stability condition ($\ref{vac-stab}$) that the $\lambda^2S^2_{min}$ term is negligible compared to $-\lambda\mu^2$ and the critical temperature ($\ref{T-cr}$) is practicaly unaffected. At this temperature the $q$ field becomes tachyonic and the global minimum of the potential moves from $q=0$ to $q= \sqrt{\frac{\mu^2}{\lambda}}$ which is the supersymmetry preserving minimum.

A second important remark is that the coupled $S$-$q$ system makes the high temperature minimum $S_{min}$ unstable (saddle point) for temperatures lower than the critical one. However, we saw that at $T=0$ there is a metastable minimum at the $S$-direction (\ref{S-min}). The temperature at which the unstable $S$-direction ($q$=0) minimum becomes a metastable one is given by the condition:
\begin{equation}
\frac{\partial^2 V(S,q)}{\partial q^2} \geq 0 \ \  \text{at} \ \ q=0, \  S=S_{min}(T) \  \text{and} \ T<T_{cr}.
\end{equation}      
First of all, we can see from ($\ref{fullT-Kitano}$) that 
\begin{equation}
\frac{\partial V(S,q)}{\partial q} = 0 \ \  \text{at} \ \ q=0.
\end{equation}      
Therefore the $V(S=S_{min},q=0)$ is an extremum of the potential. When $\partial^2 V(S=S_{min},q=0)/ \partial q^2<0$ it is a saddle point. When $\partial^2 V(S=S_{min},q=0)/ \partial q^2>0$, $S_{min}$ becomes a stable local minimum. For $T<T_{cr}$ and $\lambda>\mu^2M^2_P/\Lambda^4$ we obtain from (\ref{fullT-Kitano})
\begin{equation} \label{V''}
\frac{\partial^2 V(S=S_{min}(T),q=0)}{\partial q^2}\approx -4\lambda\mu^2+ 4\lambda^2S^2_{min}(T)
-\frac{T}{12\pi}\left[6\lambda^2\left(2\lambda\mu^2+2\lambda^2S^2_{min}(T)\right)^{1/2}\right]. 
\end{equation}      
It is easy to check that the linear  temperature correction can be safely neglected. Therefore
\begin{equation} 
\frac{\partial^2 V(S_{min}(T),q=0)}{\partial q^2}\geq0 \Rightarrow \lambda\mu^2\leq \lambda^2S^2_{min}(T)
\end{equation}      
and 
\begin{equation} \label{SminS}
S^2_{min}(T_S)\simeq\frac{\mu^2}{\lambda},
\end{equation}      
where $T_S$ the temperature at which the $S$ minimum becomes locally stable.
Equating the last relation and the equation (\ref{S-min0}) we can find the temperatue $T_S$:
\begin{equation} \label{S-minT}
S_{min}(T)=\frac{4\frac{c}{M^2_P}\mu^2-\frac{2\mu^2c}{3\Lambda^2 M^2_P}T^2}{8\frac{\mu^4}{\Lambda^2}+\frac{1}{2}\lambda^2T^2}\simeq \frac{\mu}{\sqrt{\lambda}} 
\end{equation}
which gives
\begin{equation} \label{TS-1}
T^2_S \sim \frac{c\mu}{M^2_P \sqrt{\lambda^3}}.
\end{equation}
This is a temperature typically a few orders of magnitude higher than  the tree level mass of $S$, $m_S=\mu^2/ \Lambda$, hence the high temperature expansion is valid. Below this temperature  the metastable susy breaking vacuum forms and the temperature dependent terms start becoming negligible.
Let's note that $T_S$  vanishes in the limit $M_{P}\rightarrow \infty$ which makes sense since the susy breaking vacuum disappears in this limit.
\\
\\ 
To summarize, the study of the evolution of the thermal averages of the fields (minima of the potential) from a phase of high temperature thermal equilibrium towards the zero temperature potential seems to disfavour the simple model of gravitational gauge mediation (see also \cite{Craig:2006kx}):  the susy breaking metastable vacuum is not reached if i) the univese has experienced at  high temperature  a hot thermal phase in which the hidden/messenger sector  fields ($S,q,\bar{q}$) were part of the interacting plasma  and ii) that phase sets the thermal initial conditions for the evolution of these fields, which means that they aren't displaced from the symmetric thermal minimum of the potential. 

\begin{figure} 
\textbf{\,\,\,\,\,\,\,\,\,\,\,\,\,\,\, T=0 \,\,\,\,\,\,\,\,\,\,\,\,\,\,\,\,\,\,\,\;\;\;\;\;\;\;\;\;\;\;\;\;\;\;\;\;\;\;\;\;\,\,\,\,\,\,\,\,\,\,\,\,\,\;\;\;\;\;\;\;\;\;\;\;  T$>$T$_{cr}$}
\centering
\begin{tabular}{ccc}
%{(a)} \includegraphics [scale=0.5, angle=0]{Sq1.eps} & {(b)} \includegraphics [scale=0.5, angle=0]{Sq2.eps}  &
%{(c)} \includegraphics [scale=0.5, angle=0] {Sq3.eps}  \\

{(a)} \includegraphics [scale=0.5, angle=0]{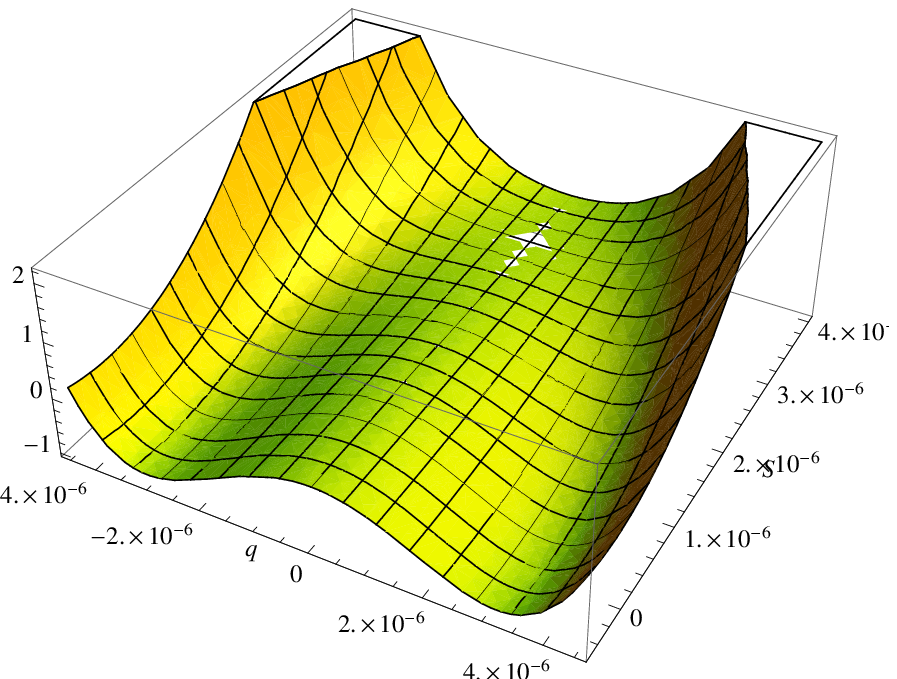} &
{(b)} \includegraphics [scale=0.5, angle=0]{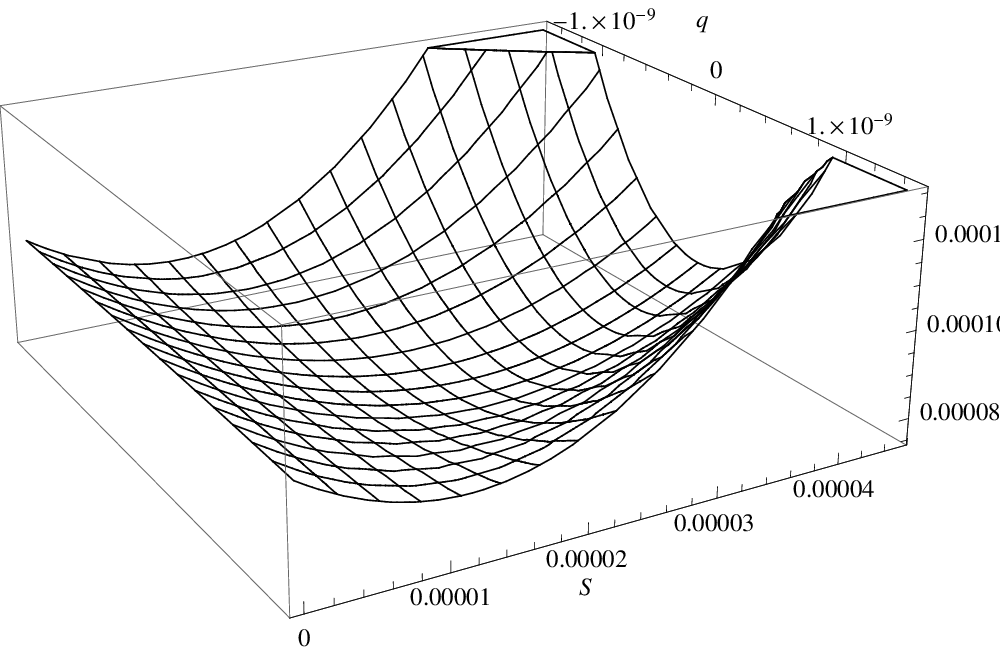}  &
{(c)} \includegraphics [scale=0.5, angle=0] {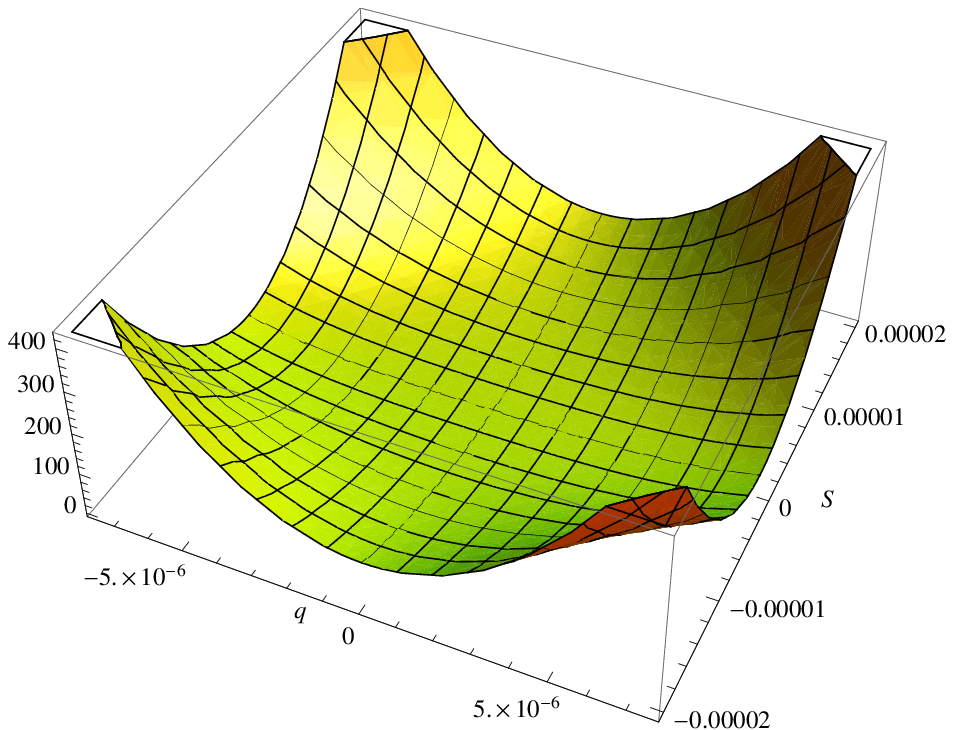}  \\
\end{tabular}
\caption{The zero temperature and finite temperature vacuum structure of the theory (\ref{KitanoK})-(\ref{KitanoW}) for $\Lambda=10^{-2}, \mu=10^{-9}$ and $\lambda=10^{-7}$. Real values for the fields are assumed. The plot in the left panel (a) depicts the supersymmetry preserving global minima at $q^2=\mu/\sqrt{\lambda}=(10^{-5.5})^2$. The shallower supersymmetry breaking metastable minimum lies further away along the $S$-direction at $S=\sqrt{3}/6\Lambda^2 \simeq 0.3 \times 10^{-4}$ and can be seen in the middle panel (b). The right panel, (c), depicts the shape of the finite temperature potential for $T>T_{cr}$ without taking into account the shift due to the ${\cal O}(T^4)$ term. The $S$, $q$ axes are scaled in Planck units while the $V(S,q)$ is given in units of $\mu^4$.} 
\end{figure}
\section{Metastable supersymmetry breaking vacua with generalized K\"ahler potential} \normalsize

We assume that at tree level the fields have canonical K\"ahler potential and that their interactions are described by the most general superpotential consistent with the $U(1)$ $R$-symmetry:
\begin{equation} \label{W-gen}
W=\mu^2 S + \tilde{\phi_i}(m_{ij}+\lambda_{ij}S)\phi_j.
\end{equation}
Interactions of $S$ with $\tilde{\phi_i}$ and $\phi_j$ induce perturbative quantum corrections in the classic theory ($\ref{W-gen}$). The leading contribution of these corrections is the one-loop Coleman-Weinberg potential for $S$ \cite{Intriligator:2006dd}, 
\begin{equation}
V_{CW}=\frac{1}{64\pi^2}Tr(-1)^F {\cal M}^4 \text{log}\frac{{\cal M}^2}{Q^2}
\end{equation}
where ${\cal M}_{ij} = m_{ij}+\lambda_{ij} S $ and $Q$ the UV cut-off of the theory.
This can be approximately accounted for by introducing a correction to the K\"ahler potential
\begin{equation} \label{K-eff}
\delta K=-\frac{1}{16\pi^2}Tr\left[{\cal M}^\dagger{\cal M}\log \left( \frac{{\cal M}^\dagger {\cal M}}{Q^2} \right)\right]
\end{equation}
where ${\cal M}_{ij} = m_{ij}+\lambda_{ij} S $ and $Q$ is the UV cut-off of the theory. The $R$-symmetry present in ($\ref{W-gen}$) implies that the $R$ charge assignment for $S$ is $R(S)=2$, and guarantees that to leading order around $S=0$ , the effective potential takes the form $V_{CW}=V_0\pm m^2_S|S|^2+ {\cal O}(|S|^4)$ \cite{Shih:2007av}. Therefore, the correction to the K\"ahler potential ($\ref{K-eff}$) is a function only of $|S|^2$ and can be expanded in powers of $|S|^2$ 
with  the dimensionful parameters $g_{2l}$:
\begin{equation} \label{K-eff3}
\delta K=\sum_{l\geq2}g_{2l}|S|^{2l}=-g_4|S|^4-g_6|S|^6-...
\end{equation}
The $g_2$ term simply rescales the canonical term in the K\"ahler potential. Hence, starting from the the superpotential ($\ref{W-gen}$) of a generalized O'R model we can integrate out the heavy chiral superfields $\phi_i$ ending up with an effective low energy superpotential $\delta W_{low}= \mu^2 S$
and a K\"ahler potential with the correction ($\ref{K-eff3}$). Actually, we integrate the supersymmetric rafertons out  for simplicity - alternatively we could keep them in the low energy Lagrangian together with the 
1-loop correction they generate. It turns out that under rather general conditions  the vacuum found in the full theory coincides to a good accuracy with the 
vacuum found in the simpler model with decoupled rafertons. Again, we neglect the gauge bosons, assuming that the nonstandard ones are very heavy, 
and noticing that the SM gauge boson masses do not depend explicitly on $\left\langle S \right\rangle$.  This gives in the global limit the effective potential 
\begin{equation}
 V_{eff}=(K_{effSS^{\dagger}})^{-1}|F_{S}|^{2}.
\end{equation}
Including gravity and messengers (not integrated out) we take 
\begin{equation} \label{W-low}
\delta W_{low}=\mu^2 S+\lambda S q \bar{q}+c.
\end{equation}
and  $K_{eff}=|S|^2-g_4|S|^4-g_6|S|^6$.

One obtains the supergravity effective potential for $q=\bar{q}$ at zero temperature
\begin{equation} \nonumber 
V_0=\mu^4-3\frac{c^2}{M^2_P}-2\frac{c}{M^2_P}\mu^2 (S+S^\dagger)+ 4 g_4 \mu^4 |S|^2 - 2\lambda \mu^2 |q|^2 +  4 \frac{c}{M^2_P} \mu^2 g_4 |S|^2(S+S^\dagger)
\end{equation} 
\begin{equation} \label{Vg}
  \;\;\;   + 2 \lambda^2 |S|^2 |q|^2 +4\mu^4 \frac{g_4}{M^2_P} |S|^4 + 9 \mu^4 g_6 |S|^4 + \lambda^2 |q|^4.
\end{equation}
Again, we are going to cancel the cosmological constant by assuming $c\approx \mu^2M_P / \sqrt{3}$. 
The dimensionful parameters $g_4$ and $g_6$ are of the form $g_4\sim \epsilon_4 \Lambda^{-2}$ and $g_6\sim \epsilon_6 \Lambda^{-4}$ where $\Lambda$ is the mass scale of the particles which are integrated out and $\epsilon_4, \epsilon_6$ are coefficients of an unspecified sign. We absorb the scale into the coefficients and write them as $|g_4|\equiv 1/\Lambda^2_1$ and $|g_6| \equiv1/\Lambda^4_2$. The model discussed  earlier in this paper corresponds to $g_4=1/\Lambda^2_1= 1/ {\Lambda^2}$ and $\Lambda_2\rightarrow\infty$. The interesting observation lies in the fact that the parameters $g_4$ and $g_6$ can be positive or negative. When the $g_4$ is positive the $g_6$ correction is negligible and we obtain the minimum already known from the previous section
\begin{equation} \label{min1}
\left\langle  S \right\rangle=\frac{\sqrt{3} \Lambda^2_1}{6 M_P}, \;  \; \; g_4>0.
\end{equation}
However, $g_4$ can be negative. We are going to keep $g_6$ positive in this case to ensure the existence of a minimum. The higher order corrections are of course considered to be $g_{2l}|S|^{2l}<|S|^2$. In terms of $\Lambda_{1,2}$ this means that our theory is valid in the regime $|S|<\Lambda_1$ and $|S|<\Lambda_2$. A simple realization of a model that leads to such  a situation has been given in \cite{Lalak:2008bc}. 

For negative $g_4$, $g_4=-|g_4|=-1/\Lambda^2_1$ the K\"ahler potential reads
\begin{equation}
K_{eff}=|S|^2+\frac{|S|^4}{\Lambda^2_1}-\frac{|S|^6}{\Lambda^4_2}
\end{equation}
and we can select the following two cases:
\begin{itemize}
	\item $S|g_4|>1/M_P \Rightarrow S>\Lambda^2_1/M_P$,
\end{itemize}
which means that the ${\cal O}(S)$ is subdominant in ($\ref{Vg}$) and leads to the minimum
\begin{equation} \label{min2}
|\left\langle S \right\rangle|^2 \sim \frac{|g_4|}{g_6} = \frac{\Lambda^4_2}{\Lambda^2_1}. 
\end{equation}
The condition $g_{2l}|S|^{2l}<|S|^2$ in this minimum implies that $\Lambda_2<\Lambda_1$. This minimum doesn't have any dependence on the $M_{P}$ which means that it survives  in the global susy limit. This can also be seen that omitting the ${\cal O}(S)$ in ($\ref{Vg}$), which contains the dimensionful constant $c$, the potential is to leading order the global susy one. The condition $S|g_4|>1/M_P$ in this minimum
translates into $\Lambda^3_1/M_P<\Lambda^2_2$ and one arrives at
\begin{equation}
\frac{\Lambda^{3/2}_1}{M^{1/2}_P}<\Lambda_2<\Lambda_1.
\end{equation}
The stability of the above vacuum gives us more constraints on the parameters. Asking for stability in the q-direction, i.e positive determinant of the $q$ mass matrix, one finds the condition 
\begin{equation} \label{vac-stab-2}
|S|^2>\frac{\mu^2}{\lambda}\Rightarrow |S|^2\sim \frac{\Lambda^4_2}{\Lambda^2_1} >\frac{\mu^2}{\lambda}.
\end{equation}
Loop corrections coming from the messengers are irrelevant if $\lambda^2 N_q/(4\pi)^2<(\Lambda_2/\Lambda_1)^4$. Thus, approximatelly, we take 
\begin{equation} \label{vac-stab-qua2}
\lambda<\left(\frac{\Lambda_2}{\Lambda_1}\right)^2.
\end{equation}
We note that contrary to (\ref{vac-stab-qua}) the above bound on the coupling $\lambda$ is not $M_{P}$ suppressed, and this is  as it should be for susy breaking minima that survive in the global susy limit. Gaugino massses of the order of $1\, {\rm TeV}$ relate the value of $\langle S \rangle$  to the $\mu^2$ as follows
\begin{equation}
\mu^2 = \left(\frac{\alpha}{4\pi} \right)^{-1} m_{\text{gaugino}} \left\langle S\right\rangle \simeq 10^{-14} \frac{\Lambda^2_2}{\Lambda_1}M_P.
\end{equation}
Combining the above constraints on $\lambda$ we find that the vacuum is metastable for  $\Lambda_2>10^{-7} \left (\frac{\Lambda_{1}^{3/2}M^{1/2}_P}{\Lambda_2} \right )$.
\begin{itemize}
	\item The second case is $S|g_4|<1/M_P \Rightarrow S<\Lambda^2_1/M_P$,
\end{itemize}
which leads to the minimum 
\begin{equation} \label{min3}
\left\langle S \right\rangle ^3 \sim \frac{1}{9}\frac{c}{\mu^2 M^2_P}\frac{1}{g_6}  \simeq  \frac{\Lambda^4_2}{M_P}.
\end{equation}
The condition $S|g_4|<1/M_P$ is fulfiled for 
\begin{equation}
\Lambda_1 \left( \frac{\Lambda_1}{M_P} \right)^{1/2} > \Lambda_2 
\end{equation}
The special limit $g_4\approx 0$ belongs to this domain. Stability of this vacuum in the q-direction implies
\begin{equation}
|S|^2>\frac{\mu^2}{\lambda}\Rightarrow S^2\sim \frac{\Lambda^{8/3}_2}{M^{2/3}_P} >\frac{\mu^2}{\lambda}
\end{equation}
and the loop correction given by messengers is irrelevant  if
\begin{equation} \label{vac-stab-qua3}
\lambda<\left(\frac{\Lambda}{M_P}\right)^{2/3}
\end{equation} 
for $\lambda^2 N_q/(4\pi)^2 \sim \lambda^2$, in accordance with (\ref{vac-stab-qua}). ${\cal O}(TeV)$ gaugino masses give us in this vacuum the relation
\begin{equation}
\mu^2  \simeq 10^{-14} \Lambda^{4/3}_2 M^{2/3}_P.
\end{equation}
Hence, here a gravitational stabilization of the vacuum is possible for $\Lambda_2>10^{-7}M_P$ which is a less stringent bound on the cut off scale compared to (\ref{Lambda}). The minimum ($\ref{min3}$), as the ($\ref{min1}$), disapears if we neglect gravity. As $M_{P}\rightarrow \infty$ the susy breaking minima ($\ref{min1}$), ($\ref{min3}$) enter the domain $|S|<\mu/\sqrt{\lambda}$ and they become tachyonic in the $q$-direction. In the absence of messengers these $M_{P}$ suppressed minima could exist even in the global susy limit where the minimum would be at $S=0$ preserving also the $R$-symmetry. However, in the presence of messengers, the gravity allows the existence of these metastable vacua where both susy and $U(1)_R$ break down. The zero temperature constraints of all the cases are assembled and presented in the Table 1.

\subsection{Temperature corrections in models  with generalized K\"ahler potential} \normalsize

If we retain only the term ${\cal O}(T^2)$ and with the tree level potential $V_0$ given by ($\ref{Vg}$), the effective potential at finite temperature takes the form
\begin{equation}
V= V_0 
+\frac{T^2}{12}\left(4g_4\mu^4+4\frac{c}{M^2_P}\mu^2g_4(S+S^\dagger)+3\lambda^2|S|^2+36\mu^4g_6|S|^2 + 6 \lambda^2 |q|^2 \right) + {\cal O}(T^4).
\end{equation}
We see that the effective potential for the $q$ field coincides with the one studied previously. There is a critical temperature $T_{cr}=2\mu/\sqrt{\lambda}$ at which a nearly second order phase transition takes place from the high temperature minimum $q=0$ to the tree level one $q^2=\mu^2/\lambda$. On the other hand, the evolution for the $S$ field changes and we study it in the following.
For $g_4<0$ we substitute $g_4=- 1/\Lambda^2_1$ and $g_6=1/\Lambda^4_2$. Near the origin, for $q=0$, the potential along the $S$-direction reads
\begin{equation} \label{V-S-T-2}
V^S=-2 \frac{c}{M^2_P}\mu^2 (S+S^\dagger)-4\frac{\mu^4}{\Lambda^2_1}|S|^2+9\frac{\mu^4}{\Lambda^4_2}|S|^4 +\frac{T^2}{12}\left(-4\frac{c\mu^2}{\Lambda^2_1 M^2_P}(S+S^\dagger)+3\lambda^2|S|^2+36\frac{\mu^4}{\Lambda^4_2}|S|^2 \right),
\end{equation}
up to terms which do not depend on $S$ and are quartic in $T$. The $\lambda^2S^2$ is the ${\cal O}(S^2)$ term that dominates in the parenthesis above since the tree level stability condition of the metastable vacuum (\ref{vac-stab-2}) implies
$\lambda^2>\mu^4/\Lambda^4_2$. We also took into account $\Lambda_1>\Lambda_2$ condition necessary for the K\"ahler potential to be well defined.
At high temperatures $T>\Lambda_1$ that the tree level terms are completely negligible the minimum is close to the origin at $S \sim c\mu^2/\lambda^2\Lambda^2_1 M^2_P$. For $T<\Lambda_1$ the ${\cal O}(ST^2)$ can be omitted in favour of the ${\cal O}(S)$ term. As the temperature decreases the $S$-minimum moves away from the origin we, again,  distinguish two cases:
\begin{itemize}
	\item $\Lambda_2> \Lambda^{3/2}_1/M^{1/2}_P \Rightarrow S(T)>\Lambda^2_1/M_P$
\end{itemize}
The temperature $T_S$ at which the minimum exits the unstable tachyonic region $|S|<\mu/\sqrt{\lambda}$ and becomes metastable is found to be 
\begin{equation} \label{T-S-2}
T_S \sim \frac{\mu^2}{\lambda}\frac{1}{\Lambda_1} .
\end{equation}
At $T_S$ the minimum is of the order of the tree level one $S^2_{min}\sim \Lambda^4_2/\Lambda^2_1$. A fast way one to find $T_S$ is to note that the above result ($\ref{T-S-2}$) would be exact in the case of absence of the gravity terms ${\cal O}(S)$ and ${\cal O}(T^2S)$ from ($\ref{V-S-T-2}$). Then the effective mass squared for $S$ would be $m^2_{S,eff}=-8\mu^4/\Lambda^2_1+T^2\lambda^2$/2 and at that temperature the $S$-minimum would move by a second order phase transition to non-zero values. This would correspond to a spontaneous $U(1)$-$R$ symmetry breaking simultaneously with the supersymmetry breaking; for another example of spontaneous $U(1)$-$R$ symmetry breaking of O'Raifeartaigh models \cite{Shih:2007av} at finite temperature, see \cite{Moreno:2009nk}. It is easy to check that the $T_S$ ($\ref{T-S-2}$) is less than the $T_{cr}=2\mu/\sqrt{\lambda}$ at which a second order phase transition to the $q$ direction takes place. A way to understand this qualitatively is that the direction that opens first, i.e. becomes tachyonic, is the one towards the minima closest to the origin. The susy breaking minima are shallower than the susy preserving ones and can be locally stable only if they are further than the susy minima (condition (\ref{vac-stab-2})).
\begin{itemize}
	\item $\Lambda_2< \Lambda^{3/2}_1/M^{1/2}_P \Rightarrow S(T)<\Lambda^2_1/M_P$
\end{itemize}
Here, the temperature at which the minimum exits the tachyonic region $|S|<\mu/\sqrt{\lambda}$ and becomes metastable is  
\begin{equation} \label{T-S-3}
T^2_S \sim \frac{c\mu}{M^2_P\sqrt{\lambda^3}}
\end{equation}
This can be seen from the fact that for $S(T)<\Lambda^2_1/M_P$ the ${\cal O}(S)$ term dominates over the ${\cal O}(S^2)$ in ($\ref{T-S-2}$). At this temperature the minimum is to a good approximation the zero temperature (tree level) one $S^3_{min} \sim \Lambda^4_2/M_P$. We see that ($\ref{T-S-3}$) coincides with ($\ref{TS-1}$) as actually expected from the similarities of the two models. We again remark that ($\ref{T-S-3}$) is $M_{P}$ suppressed which means that it disappears on the global susy limit together with the susy breaking minimum.

To summarize, we have discussed models that possess susy preserving vacua close to the origin and metastable susy breaking vacuua at VEVs defined by a power of an intermediate scale $\Lambda$ (or $\Lambda_1,\Lambda_2$) which is the cut-off for  these theories. We saw that at high temperatures the field is trapped near the origin. As the universe cools down, at the temperature $T_{cr}=2\mu/\sqrt{\lambda} > T_S$, there is a second-order phase transition. The origin becomes unstable, since the $q$-direction becomes tachyonic, and the fields land in the supersymmetric global minimum. The small non-zero expectation value along the $S$-direction cannot block the transition to the supersymmetric vacuum. The conclusion seems to be that the susy-breaking metastable vacuum is not realized in the early universe. 

\section{Evolution of spurion and messengers  during inflation and reheating} \label{infla-reh}\normalsize

\subsection{Displaced initial conditions for  spurion and  messengers}
In the previous sections we have assumed that susy breaking sector is thermalized after the reheating of the universe without considering how natural this assumption is within a standard cosmological framework. During inflation the fields are expected to be displaced from the zero temperature minimum \footnote{Asymmetric initial conditions for the scalar fields is nothing novel even outside the inflationary framework; see \cite{Langacker:1980kd} for an early example.}. This displacement can be due to gravitationally produced fluctuations or due to a coupling of a field to the inflaton $I$ - in the superpotential or via corrections to the K\"ahler potential. We will express this initial dispacement as $S_\text{INF}$ denoting its connection to the inflationary dynamics. If the field $S$ has a mass  which is small compared to the inflationary Hubble scale $H_I$ then its fluctuations are of the order of $H_I$, \cite{Felder:1999wt}, which is understood as providing the initial value $S_\text{INF}$ for the field S. Based on the COBE normalization of inflationary fluctuations one can estimate $ S_\text{INF} \sim H_{I} \sim 10^{-5} M_P$. 
Massive fields can be given a large expectation value via a direct coupling  to the inflaton. 
The $F$-component of the inflaton, $F_I$, dominates the energy density during inflation: $|F_I|^2=3H^2 M^2_P$.  For instance, the singlet $S$ may couple to the inflaton sector in the superpotential via the term 
\begin{equation}
W=W_I \left (1-\xi \frac{ Q_I }{M^2_P}S+...\right )
\end{equation}
where $W_I$ denotes the superpotential of an inflaton sector, $\xi$ a coefficient of ${\cal O}(1)$ and $Q_I$ an energy scale connected to the inflationary dynamics. A possibility is that $Q_I = I-I_0$ is the displacement of the inflaton away from its post-inflationary minimum, computed at the end of inflation (see also \cite{Ibe:2006rc}). The scalar potential during inflation for the field $S$  takes the form 
\begin{equation} \label{V-I}
V\simeq 3H^2 \left (|S|^2- {\cal O} \left(\frac{|S|^4}{\tilde\Lambda^2}\right)\right) + 3H^2 M^2_P \left|1-\xi\frac{ Q_I}{M^2_P} S\right|^2.
\end{equation}
The minimum of this potential is $S={\cal O}(Q_I)$. 
If $Q_I$ is interpreted as the displacement $I-I_0$ then small field inflationary models  should be assumed in order that the K\"ahler potential of our theories stays  positive definite. 
As for gauge non-singlet messengers the more direct way to displace them in a controllable way is to use interactions in the K\"ahler potential \cite{Dine:1995uk}, for instance $\delta K = - q^\dagger q I^\dagger I/ M^2_P $, which induces a negative effective mass squared of the order of $H^2$ and this way pushes the field away from the origin. 

 After inflation the inflaton is supposed to oscillate about the minimum of its potential and via preheating/reheating \cite{Kofman:1994rk} decays and a radiation dominated phase takes over. However, until its decay  the potential ($\ref{V-I}$) is approximately valid. The Hubble scale decreases, and the fields trace the minimum of the potential  moving towards the origin.  
Qualitatively, in the region $S<\Lambda$ according to (\ref{Kitano-tree}) and (\ref{V-I}) the scalar potential for $S$ can be described by 
\begin{equation}
V \sim 4\frac{\mu^4}{\Lambda^2}|S|^2-2\frac{c}{M^2_P}\mu^2(S+S^{\dagger})+3 H^2\left(|S|^2-Q_I(S+S^\dagger)\right)
\end{equation}
which can be rewritten as 
\begin{equation} \label{V-I2}
V\sim \frac{1}{2}(8 \frac{\mu^4}{\Lambda^2}+6H^2)|S|^2-(2\frac{c}{M^2_P}\mu^2+3H^2Q_I)(S+S^\dagger).
\end{equation}
The minimum of this potential for $S$ lies at
\begin{equation} \label{full_min}
S_{min}\equiv {\rm Re} (S)=\frac{4c\mu^2/M^2_P+6H^2Q_I}{8 \frac{\mu^4}{\Lambda^2}+6H^2}, \,\,\,\,\,\, {\rm Im} (S)=0.
\end{equation}
Right at the end of inflation the Hubble parameter is the dominant one in (\ref{V-I2}) and the $\left\langle S \right\rangle$ is well localized at the minimum $Q_I$. If $Q_I$ is a constant energy scale then the field will remain frozen  until the tree level parameters become relevant.
In case $Q_I =I -I_0$ then $Q_I$ decreases together with the amplitude of the inflaton coherent oscillations. 
Under the assumption that the effective mass of $S$ is large enough (larger than $H$) the position of $S$ will follow the minimum of the potential to settle down in the metastable supersymmetry breaking zero-temperature minimum. Hence, the spurion at the time of reheating has vev $S_\text{rh}\sim \Lambda^2/M_P$.
The answer to the question of whether such a tracing of the minimum  does indeed take place depends on details of the model.

In conclusion, the initial values of the spurion right after inflation, $S_\text{INF}$, and at at the time of inflaton decay, $S_\text{rh}$, are very model dependent. Nevertheless, $S_\text{INF}$ is expected to be related to the mass scale of inflation. 

\subsection{Thermalization}

At some point the inflaton decays completely and the universe becomes reheated. After reheating $I=I_0$ and there are no Hubble induced terms in the potential of $S$, $q$ and $\bar{q}$. The value of the spurion at that time is denoted as $S_{rh}$. If the interaction rate of these fields with the thermal plasma is larger than the expansion rate $H$, the fields will  thermalize and the potential will be corrected by temperature dependent terms. Otherwise, their potential will be the zero temperature one, i.e. the tree level potential enhanced by  Coleman-Weinberg loop corrections. 

The messengers $q$ and $\bar{q}$ are coupled to the MSSM plasma via the SM gauge forces and they can achieve thermal equilibrium. If the reheating temperature is higher than $T_q=m_q/20$ where $m_q \approx \lambda \left\langle S \right\rangle$ then they get thermalized and stabilized at the origin. If not, one can still expect a thermal mass for messengers due to 
%it is still expected to have vevs $\left\langle q \right\rangle=\left\langle \bar{q}\right\rangle=0$ 
due to thermalized MSSM gauge bosons. The messengers SM gauge numbers couple to the gauge bosons as follows: 
\begin{equation} \label{thmass}
V \subset g^2(|q|^2+|\bar{q}^2|)\left\langle A_\mu A^\mu\right\rangle_T. 
\end{equation}
This induces a thermal mass of the order of $g\,T$,  large enough to push  the vevs of the messengers towards  $\left\langle q \right\rangle=\left\langle \bar{q}\right\rangle=0$  even for $T_\text{rh}<T_q$.

On the other hand, the spurion  $S$ is coupled to MSSM degrees of freedom via loop diagrams with the messengers as the heavy fields in the loops.  For $T>m_q$ the  thermally averaged cross section for 2-2 processes  with two spurions is of the order of 
\begin{equation} 
\Gamma_{int}=\left\langle \sigma v n\right\rangle_{T}\sim \frac{\lambda^4 \alpha^2 }{16 \pi^2} T
\end{equation}
($\alpha$ corresponds to the SM fine structure constant) and the equilibrium in a radiation dominated universe, $\Gamma\geq H=\sqrt{g_*} \, T^2/M_P$, may be achieved below the temperature 
\begin{equation} \label{Teq}
T_{eq} \sim \frac{\lambda^4 \alpha^2 }{16 \pi^2 \sqrt{g_*}}M_P={\cal O}(10^{-3})\lambda^4 \alpha^2 M_P.
\end{equation}
When messengers become  non-relativistic, i.e. for  $T< m_q$, the thermalization could also be achieved and the relevant  averaged cross section becomes $ \left\langle \sigma v n \right\rangle_{T}\sim \alpha^2 \lambda^4T^5/ m^4_q$ .  The requirement that this interaction rate is larger than the expansion rate gives a lower bound on the temperatures at which  $S$ can be  thermalized. If the coupling $\lambda$ is small enough then the window of temperatures where the spurion  thermalizes can be closed. Actually, the zero temperature constraints on the coupling $\lambda$ (\ref{vac-stab-qua}), (\ref{vac-stab-qua2}) and (\ref{vac-stab-qua3}) don't allow the $S$ field to thermalize. Even when $T>m_q$, i.e. when the messengers running in the loops are light compared to the temperature, the spurion $S$ is out of equilibrium.

However, when $T_\text{rh}>T_q$, the messengers get thermalized and they can contribute thermal corrections to the potential of the spurion. 
One can see this if one takes into account that thermal averaging of the  term $\sim \lambda^2 |S|^2 |q|^2$ in the tree level scalar potential leads to 
\begin{equation} \label{S-thermal}
\lambda^2 |S|^2 \langle |q|^2 \rangle_T \sim \lambda^2 |S|^2 T^2,
\end{equation}
for a thermal distribution of messengers. Once $T<T_q$ the evolution of the spurion is governed by the zero temperature potential.

It is interesting to note that thermalized MSSM degrees of freedom may alter the value of  the critical temperature. The exact modification of the critical temperature depends on the Lagrangian that describes the interactions of the messengers with the observable sector and it is generally model dependent. The study of the coupled system of hidden, messenger and observable sectors lies beyond the scope of this paper and shall be discussed separately. 

In conclusion, the messengers being charged under the Standard Model gauge group obtain thermal masses which help to localize them at the origin of the field space. When the temperature is higher than their tree level mass $m_q \simeq \lambda \left\langle S \right\rangle$ they induce a thermal mass for the spurion $S$ according to (\ref{S-thermal}). This thermal mass may also drive the $S$ field to the origin. On the contrary, when the temperature is lower than $m_q/20$ the thermal excitations of messengers are Boltzmann suppressed hence the thermal induced mass of the spurion is negligible. 
%In the following we will show that for sufficietly small coupling the thermalized messengers don't rule out the selection of the metastable vacuum. 

\section{Conditions for selection of the susy breaking vacuum}

The condition which controls the selection of a susy breaking vacuum is 
\begin{equation} \label{condition1}
\mu/\sqrt{\lambda}< |S|<{\Lambda}. 
\end{equation}	
This is the prospective basin of attraction of the susy-breaking minimum. It is bounded from above by the condition that the quantum corrected kinetic energy stays positive definite. The lower bound is the tachyonic region about the origin. This  condition can be broken into two: i) $S$ at the end of inflation should have a vev in  the regime of our effective theory, i.e. $|S_\text{INF}|<\Lambda$. 
%If this condition is violated then a UV theory must be considered instead of the low energy one. 
ii) $S$ during reheating, or during a non-thermal phase, must obey $|S|>\mu/\sqrt{\lambda}$. In the case of a generalized K\"ahler potential the cut-off scale is $\Lambda_2$ and the condition reads: $\mu/\sqrt{\lambda} < |S|<{\Lambda_2}$.
\\
\\
Unless the above condition is fulfilled the system lands either in a susy preserving vacuum or in the region of  large field values where our IR effective theory is not valid. For the selection of the susy breaking vacuum messengers must have a vev $\left\langle q \right\rangle=0$. Otherwise, the spurion has a tree level mass contribution $\lambda \left\langle q \right\rangle$ and can be attracted to the origin. As we have noted  in the previous section, we expect the messengers to have a vev $ \left\langle q \right\rangle=0$ thanks to the thermal mass that messengers  receive from the  MSSM gauge bosons even in the case they are not thermalized. 
%Furthermore, since large vevs for messengers result in a hard breaking of SM gauge symmetries a trapping effect due to the heavy gauge bosons %should enhance the attraction of messengers. 

However, in the case the messengers are thermalized the spurion also receives a thermal mass. This mass drives the spurion to the origin. A simple  way out would be to impose the condition  $T_\text{rh} <T_q=m_q/20$. But, this solution doesn't guarantee that the messengers are driven to $\langle q \rangle = \langle \bar q \rangle =0$. What helps is the observation, that messengers even if not thermalized, receive a thermal mass (\ref{thmass}), which 
does prefer $\langle q \rangle = \langle \bar q \rangle =0$.
%Due to this fact, it was usually asked the extra condition that the reheating temperature \footnote{Reheating coming from the inflaton decay or the decay %of any other field that causes late entropy production, including the spurion field itself. In case the spurion reheats the universe this condition is naturally %fulfilled, as the reheating of the weakly coupled field cannot be efficient.}  is lower than $T_\text{rh} <T_q=m_q/20$. 
%Here, we argue that actually thermal mass of the messengers helps to select vanishing vevs for the messengers. 
%the opposite: thermalized messengers do \itshape not \normalfont necessarily exclude the selection of the susy breaking vacuum. 

In fact, the more general condition which makes it likely  that the metastable vacuum becomes actually selected is the following: 
\begin{equation} \label{Trh-suf} 
		T_\text{rh}<\text{max}\{T_q, T_S\}.
\end{equation}
We note that the temperature $T_S$ can be larger than the mass of messengers and hence they can be thermalized.  The point is again the existence of the thermal mass (\ref{thmass}) which implies that messengers are actually driven to the origin at a faster rate than the spurion which has thermal mass $\lambda T \ll g \, T$.
To summarize, the result is that the system of fields  is aligned along the $S$-direction (in the complex $S$-plane) and the tree level $q$-contribution to the mass of the $S$ field, $\lambda \left\langle q \right\rangle$, vanishes. 
%\footnote{Taking into account the couplings of messengers to MSSM degrees of freedom enhances the messengers' thermal mass. This means that %messengers are actually driven to the origin much faster than the spurion which has thermal mass $\lambda T \ll T$. }. 
%As has been said before, thermal excitations of messenger fields may  also induce a thermal mass for the spurion $S$. 
The upper bound on the reheating temperature (\ref{Trh-suf}) guarantees that the susy breaking vacuum in the $S$ direction has formed: it is locally stable. Thus, the field can land in the susy breaking vacuum. In particular, below the temperature $T=\mu^2/(\lambda\Lambda)<T_S$ the relevant potential is approximately the zero temperature one (\ref{V-zero})
\begin{equation}
V(S)\simeq \mu^4-\frac{3c^2}{M^2_P}-2\frac{c}{M^2_P}\mu^2 (S+S^\dagger)+4\mu^4 \frac{|S|^2}{{\Lambda}^2} + \frac{\lambda^2 N_q}{(4\pi)^2} \mu^4 \log \frac{S^\dagger S}{Q^2} +{\cal O}(T).
\end{equation}
The logarithmic term is important near $S=0$ but its effects are negligible for small values of coupling $\lambda\ll \Lambda$. As shown in \cite{Ibe:2006rc}, where a non-thermal evolution of the system of fields was considered, the $S$ field feels at most points of the complex $S$-plane a much stronger force towards the supersymmetry breaking vacuum than towards the supersymmetric one. It has been shown numerically that for initial conditions Re$(S)$=Im$(S)=\Lambda$ the $S$ field settles into the supersymmetry breaking minimum. If we additionally want the energy stored in the oscillations of the $S$ field not to dominate the energy density of the universe then the spurion at the time of reheating should be localized around the metastable minimum. This could be realized via a possible tracing of the minimum e.g. via the mechanism sketched in Section 6.1 and the equation (\ref{full_min}). Otherwise, it is possible that the late decay of the spurion will cause a late entropy production diluting the dark matter abundance and the baryon asymmetry.

The condition (\ref{Trh-suf}) is not a strict constraint on the reheating temperature. Actually, the smaller the coupling $\lambda$  the larger the reheating temperature can be. Also, decreasing the $\lambda$ opens the window $m_q<T_\text{rh}<T_S$, see Table 3.

An interesting observation here is that for couplings $\lambda$ so small that $\lambda<T_S/M_P$ the upper bound on the reheating temperature at (\ref{Trh-suf}) is not necessery. The reheating temperature can be arbitrary high. The reason is that due to the very small coupling $\lambda\ll 1$ the thermal mass $\lambda T$ is smaller  than the Hubble scale $H \simeq \sqrt{g_*}\,T^2/M_P$ and the field $S$ starts rolling down only for temperatures $T<T_S$. This can be seen by examining the dynamics of the spurion. In an FRW universe the homogeneous spurion field obeys the equation of motion
\begin{equation}
\ddot{S}+3H\dot{S}+ dV/dS=0.
\end{equation}
In a radiation dominated phase $H=1/(2t)\simeq \sqrt{g_*}\,T^2(t)/M_P$ and   $dV/dS\sim \mu^4/\Lambda^2 S+\lambda^2T^2S+\lambda^2q^2S$. Taking into account that the thermalized messengers are driven fast to the origin, one can see that  for $T\geq T_S$ the thermally  induced spurion mass dominates the potential and the equation of motion reads
\begin{equation}
\ddot{S}+3\sqrt{g_*}\,T^2(t)\dot{S}+ \lambda^2T^2(t)S \approx 0.
\end{equation}
The $S$ starts rolling down only after $H\sim m_S(T)$ i.e. when $\lambda \sim T/M_P$. Hence, the condition $\lambda<T_S/M_P$ means that the thermal corrections to the spurion will not drive it to the origin before the metastable susy breaking vacuum has appeared, independently  of how large the reheating temperature actually is. However, there is a price to pay: the smallest the coupling $\lambda$ the largest is the tachyonic region $|S|<\mu/\sqrt{\lambda}$ about the origin and therefore, the area of the initial values for the field $S$ which realize the metastable vacuum gets reduced. 

\section{Constraints on the susy breaking}

In this section 
we are going to review the constraints coming from the vacuum stability at zero temperature which have been already considered and combine them with cosmological constraints and constraints coming from gauge mediation  domination over gravity. 
The zero temperature vacuum stability constraints which have been already discussed in previous chapters have been summarized in Table 1.
\begin{table} \label{tabb1}
\begin{center}
$$
\begin{array}{|r|r|r|r|r|}
\hline
 K= |S|^2 \mp \frac{|S|^4}{\Lambda^2_1}-\frac{|S|^6}{\Lambda^4_2} & {\rm Metastable \; Vacuum} \left\langle S\right\rangle & \Lambda_1 \,\,\,\,\,& \Lambda_2 \,\,\,\,\,\,\,\,\,\,\,\,\,\,\,& \lambda \,\,\,\,\,\,\,\,\,\,\,\,\,\,\,\,\,\,\,\, \\
\hline 
1.\, ~(-),\, \; \, \Lambda_2 = \Lambda_1 \equiv  \Lambda & \left\langle S\right\rangle \sim \Lambda^2 \,\,\,\,\,\,\,\,\,\,\,\,\,\,\,\,\,& \Lambda > 10^{-14/3} & -\,\,\,\,\,\,\,\,\,\,\,\,\,\,\,\,\,\,\,\, & 10^{-14}\left\langle S\right\rangle^{-1}<\lambda<\Lambda \,\,\,\,\, \\
\hline
2. ~~~~~  (+), \, \; \, \Lambda^{3/2}_1<\Lambda_2  & |\left\langle S\right\rangle| \sim \frac{\Lambda^2_2}{\Lambda_1}\,\,\,\,\,\,\,\,\,\,\,\,\,\,\, & \Lambda_1>\Lambda_2 & \Lambda_2> 10^{-7} \left(\frac{\Lambda^{3/2}_1}{\Lambda_2}\right)& 10^{-14}\left\langle S\right\rangle^{-1}<\lambda< \left(\frac{\Lambda_2}{\Lambda_1}\right)^2 \\
\hline
3. ~~~~~  (+), \, \; \, \Lambda^{3/2}_1>\Lambda_2 & \left\langle S\right\rangle \sim \Lambda^{4/3}_2\,\,\,\,\,\,\,\,\,\,\,\,\,\,\, & \Lambda_1>\Lambda_2 & \Lambda_2> 10^{-7} & 10^{-14}\left\langle S\right\rangle^{-1}<\lambda< \Lambda_2^{2/3} \\
\hline
\end{array}
$$
\end{center}
\caption{\small Zero temperature vacuum stability constraints ($M_P=1$) in the three cases of the generalized K\"ahler potential. With gaugino masses $m_{\text{gaugino}} \propto \mu^2/ \left\langle S\right\rangle$ fixed at ${\cal O}(100\ \text{GeV}-1\ \text{TeV})$ the free parametrs left are the cut-off scale and the coupling $\lambda$. The first case corresponds to $\Lambda_1=\Lambda_2=\Lambda$. Since the 4th order correction has the same sign as the 6th order one, the 6th order term is negligible. In the other two cases the 6th order correction is necessery for the stabilization of the metastable vacuum. The hierarchy between $\Lambda_1$ and $\Lambda_2$ i.e. $\Lambda_1>\Lambda_2$ keeps the corrections to the K\"ahler potential under control. The lower bound on the cut-off scales $\Lambda$, $\Lambda_2$ originates from the vacuum meta-stability in the messenger sector. The upper bound on the coupling $\lambda$ renders the one-loop correction to the $S$ potential irrelevant for the stability of the vacuum and the lower bound prevents messengers from becoming tachyonic.}
\end{table}
Asking for gauge mediation domination over gravity mediation we can further constrain the cut-off parameters $\Lambda, \, \Lambda_1$ and $\Lambda_2$.  If gravity mediation contribution to squark mass squared is suppressed to ${\cal O}(1\%)$ then FCNC are sufficiently suppressed \cite{Feng:2007ke}. Hence, in the metastable vacuum we ask for 
\begin{equation} \label{grav/gau}
\frac{m_{3/2}}{m_{\text{gaugino}}}= \frac{4\pi}{\alpha\sqrt{3}}\frac{\left\langle S \right\rangle}{M_{P}}< {\cal O}(10\%).
\end{equation}
%where we reintroduced momentarily the $M_{Pl}$. 
Obviously this gives us an upper bound on the value of the spurion $S$ field. For $\alpha=0.04$ and %$M_{Pl}=1$ 
it yields
\begin{equation} \label{Smax}
\left\langle S \right\rangle \leq{\cal O}(10^{-4}-10^{-3})M_P
\end{equation}
and for the rest of the paper we take the conservative bound $\left\langle S \right\rangle \leq 10^{-4}M_P$. 

For the first case where the 6th order correction to K\"ahler potential is negligible the above constraint translates into $\Lambda\lesssim10^{-2} M_P$. For the second and third cases where the 6th order correction to K\"ahler potential is necessery we respectively have $\Lambda_2\lesssim \sqrt{10^{-4}\Lambda_1 M_P}$ and $\Lambda_2\lesssim 10^{-3} M_P$. In the second case, the $\Lambda^{3/2}_1 /M^{1/2}_P<\Lambda_2$ condition gives us a numerical upper bound on $\Lambda_2$: $\Lambda_2\lesssim \sqrt{10^{-4}\Lambda_1 M_P}<10^{-3} M_P$. We see that the stringent upper bounds on the cut-off scales apply for the case of K\"ahler potential corrected up to 6th order and especially for the second case in which susy breaking vacua survive in the global limit. 

We can additionally apply the cosmological constraints presented in the previous section. The necessery condition for selection of the metastable susy breaking vacuum is that the initial value of the spurion $S$ is smaller than the cut-off scale $\Lambda$, $\Lambda_2$ ($\Lambda_1>\Lambda_2$).
We have set this initial value right after inflation to be $ S_\text{INF}$.
We shall  assume  $|S_\text{INF}| \ll M_P$. 
 Hence, we ask for $\Lambda>|S_\text{INF}|$ and $\Lambda_2>|S_\text{INF}|$. The larger the $|S_\text{INF}|$ is the larger the cut-off scale has to be. 
 For the three types of models these constraints are presented in the Table 2.

As discussed in Section 7 there exists  an additional condition which favours the selection of the susy breaking vacuum. Namely: $T_\text{rh}<\text{max}\{T_q, T_S\}$. The temperature, at which the metastable vacuum appears, $T_S$, depends inversely on the coupling $\lambda$. Hence, the smaller the coupling is the higher the $T_S$. In other words, small coupling $\lambda$ means a small coupling to the thermal bath i.e. the tree level potential dominates over the 1-loop temperature dependent corrections even for high temperatures. On the other hand, the larger the coupling $\lambda$ the heavier the messengers are. Of course, the value of $\lambda$ is model dependent. But, it cannot be arbitrary large or arbitrary small. For example, considering the first case of gravitational stabilization, where the order 6 correction is negligible, the coupling lies in the range $10^{-14}(\Lambda/M_P)^{-2}<\lambda<\Lambda/M_P$, see Table 1. Otherwise, the vacuum is unstable either due to tachyonic direction or due to quantum correction coming from the coupling $\lambda S q\bar{q}$.
\begin{table} \label{tabb2}
\begin{center}
$$
\begin{array}{|r|r|r|r|}\hline
K=|S|^2\mp \frac{|S|^4}{\Lambda^2_1}-\frac{|S|^6}{\Lambda^4_2} & \Lambda_1 \,\,\,\,\,\,\,\,\,\, & \Lambda_2 \,\,\,\,\,\,\,\,\,\,\,\,\,\,\,\,\,\,\,\, & \lambda \,\,\,\,\,\,\,\,\,\,\,\,\,\,\, \\
\hline 
1.\, ~(-),\, \; \, \Lambda_2=\Lambda_1\equiv \Lambda   & |S_\text{INF}|<\Lambda\lesssim 10^{-2} & - \,\,\,\,\,\,\,\,\,\,\,\,\,\,\,\,\,\,\,\,& 10^{-14}\Lambda^{-2}<\lambda< \Lambda \,\,\,\,\, \\
\hline
2.~~~~~ (+), \, \; \, \Lambda^{3/2}_1<\Lambda_2  & \Lambda_1>\Lambda_2 \,\,\,\,\,\,\,\,\,\,& |S_\text{INF}| <\Lambda_2\lesssim \left(10^{-4} \Lambda_1 \right)^{1/2}& 10^{-14}(\Lambda^2_2/\Lambda_1)^{-1}<\lambda< \left(\frac{\Lambda_2}{\Lambda_1}\right)^2 \\
\hline
3. ~~~~~ (+), \, \; \, \Lambda^{3/2}_1>\Lambda_2   & \Lambda_1>\Lambda_2 \,\,\,\,\,\,\,\,\,\,& |S_\text{INF}| <\Lambda_2\lesssim 10^{-3} & 10^{-14}\Lambda_2^{-4/3}<\lambda< \Lambda_2^{2/3} \\
\hline\end{array}
$$
\end{center}
\caption{\small Combined constraints ($M_P=1$) coming from zero temperature vacuum stability, cosmological considerations necessary for susy breaking vacuum selection  and gauge mediation domination over gravity. The $S_\text{INF}$ is the initial value of the spurion field right after the end of the inflationary phase. It constrains the cut off scales from below and the requirement for gauge mediation domination over gravity constrains them from above. The larger the $S_\text{INF}$ is the more the gravity contributes to the susy breakdown mediation. The constraints on the coupling $\lambda$ are the zero temperature ones.}
\end{table}

We explained previously that the messengers can be thermalized without ruling out the selection of the metastable vacuum. This is achieved when $T_q<T_\text{rh}<T_S$. This translates in a constraint on the coupling which for large $T_S=(10^{-42}\Lambda^6/(\lambda^3 M^2_P))^{1/4}$ has to be small.  Furthermore, as mentioned in Section  7, if the coupling $\lambda$ is smaller than $T_S$ the selection of the metastable vacuum does not imply any bound on the reheating temperature. The reason is that if $\lambda<T_S/M_P$ i.e. $\lambda<10^{-6}(\Lambda/M_P)^{6/7}$ the spurion will not roll unless the temperature drops below $T_S$. 
We note that such small values of the coupling $\lambda$ are reasonable if the $S$ field is a composite operator above the scale $\Lambda$ as is often the case in dynamical supersymmetry breaking scenarios. Then $\lambda$ is suppressed by a factor of $(\Lambda/M_P)^{d(S)-1}$ where $d(S)$ the dimension of the operator $S$ above the scale $\Lambda$ \cite{Ibe:2007km}. Small values of $\lambda$ imply stabilization of the spurion at relatively large vevs, hence, the gravitino mass lying in the GeV range.

For the first case of negligible 6th order K\"ahler correction,  Table 3 demonstrates the range of values of the reheating temperature that can stabilize the spurion at the susy breaking vacuum for different values of the parameter $\lambda$.
\\
\begin{table} \label{tabb3}
\begin{center}\begin{tabular}{|r|r|r|r|r|}
\hline
$\Lambda$ & $\lambda$ & $T_q \simeq m_q/20 \,\,\,\,\,\,\,\,\,\,\,\,\,$ & $T_{S}\;\;$ & $T_\text{rh} \,\,\,\,\,\,\,\,\,\,\,\,\,\,\,\,\,\;\;$ \\
\hline
$10^{-2}$ & $10^{-3}$ & $ 10^{-8.3}<T_q< 10^{-6.3} $ & $10^{-11.25}$ &  $T_\text{rh}<T_q$\,\,\,\,\,\,\,\,\,\,\\
\hline 
$10^{-2}$ & $10^{-5}$ & $ 10^{-10.3}<T_q< 10^{-8.3} $ & $10^{-9.75}$ &  $T_\text{rh}<\text{max}\{T_q, T_S\}$\\
\hline
$10^{-2}$ & $10^{-7}$ & $ 10^{-12.3}<T_q<10^{-10.3} $ & $10^{-8.25}$ &  $T_\text{rh}<T_S$ \,\,\,\,\,\,\,\,\,\, \\
\hline
$10^{-2}$ & $10^{-8}$ & $ 10^{-13.3}<T_q<10^{-11.3} $ & $10^{-7.5}$ & $  \text{unbounded} $ \,\,\,\,\,\,\,\,\,\, \\
\hline
$10^{-2}$ & $10^{-9}$ & $ 10^{-14.3}<T_q<10^{-12.3}$ & $10^{-6.25}$  & $ \text{unbounded} $ \,\,\,\,\,\,\,\,\,\, \\
\hline
$10^{-3}$ & $10^{-4}$ & $10^{-11.3}<T_q<10^{-8.3} $ & $10^{-12}$ & $T_\text{rh}<T_q$ \,\,\,\,\,\,\,\,\,\, \\
\hline
$10^{-3}$ & $10^{-7}$ & $10^{-14.3}<T_q<10^{-11.3} $ & $10^{-9.75}$ & $T_\text{rh}<T_S$ \,\,\,\,\,\,\,\,\,\, \\
\hline
$10^{-4}$ & $10^{-5}$ & $10^{-14.3}<T_q<10^{-10.3} $ & $10^{-12.75}$ & $T_\text{rh}<\text{max}\{T_q, T_S\}$\\
\hline
\end{tabular}\end{center}
\caption{\small Some bounds on the reheating temperature that favour, for specific initial values for the spurion, the selection of the susy breaking vacuum by the system of fields.  The $\Lambda$ cannot exceed $\sim 10^{-2}$ which is the maximum value allowed by the requirement of gauge mediation domination. The reheating temperature is \itshape not \normalfont bounded for couplings $\lambda<T_S$. If not, the upper bound is either $T_S$ or $T_q$ if $T_q\simeq m_q/20>T_S$. Approximately, the maximum value of $m_q$ is $\lambda \Lambda$ and the minimum $\lambda \Lambda^2$. We used $M_P=1$ and these results are for the first case i.e.  the K\"ahler potential for the spurion is $K=|S|^2-|S|^4/\Lambda^2$.}

\end{table}

\begin{figure}[htbp]\begin{center}
\includegraphics[width=0.595\linewidth]{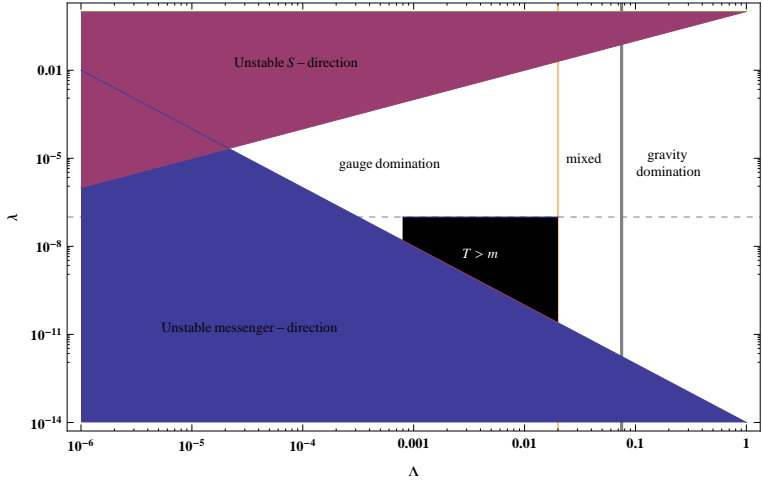}
\caption{\small The parameter space spanned by the energy scale $\Lambda\approx \sqrt{\left\langle S \right\rangle}$ (horizontal axis with $M_P=1$) and the coupling $\lambda$ (vertical axis) for the case of the K\"ahler potential $K=|S|^2-|S|^4/\Lambda^2$. The white triangle region (including the black trapezium) gives a metastable supersymmetry breaking minimum. It is divided into three parts. The part on  the right of the parameter space corresponds to $m_{3/2}>m_\text{gaugino}$ i.e. to gravity domination in the mediation of susy breaking. The part on the left corresponds to $m_{3/2}<0.1 m_\text{gaugino}$ according to (\ref{grav/gau}). The part in between gives susy breaking minima with mixed gauge-gravity mediation. The \itshape black region \normalfont is the parameter space  that allows the realization of the metastable minimum for thermalized messengers, $T_\text{rh}>T_q=m_q/20$. It could be extended to the right of the parameter space, i.e. towards larger $\Lambda$, if the requirement of gauge mediation domination is waved.}   
\end{center}
\end{figure}

\section{Including a bare messenger mass}

We generalize the superpotential ($\ref{KitanoW}$) including an explicit mass term $M$ for the messengers:
\begin{equation} \label{MNW}
W=\mu^2 S-\lambda Sq\bar{q}\pm M q\bar{q}+c.
\end{equation} 
keeping the same structure ($\ref{KitanoK}$) for the K\"ahler potential 
\begin{equation} \label{MNK}
K=S^{\dagger}S-\frac{(S^\dagger S)^2}{\Lambda^2}+ {\cal O} \left(\frac{(S^\dagger S)^3}{\Lambda^4}\right)+q^\dagger q+\bar{q}^\dagger \bar{q}.
\end{equation} 
The mass terms violates the $U(1)$ $R$-symmetry down to a $Z_2$ one. We are assuming that the  messenger mass is fixed in the fundamental theory.
This model with $\delta W= -Mq\bar{q}$ and the K\"ahler potential (\ref{MNK}) was discussed  in \cite{Murayama:2007fe} in the global susy framework. Here, we will couple it to gravity and comment on the thermal behaviour of such a model. 

In the global susy the theory the tree level potential reads
\begin{equation}
V_0=|\mu^2-\lambda q\bar{q}|^2\left(1+\frac{|S|^2}{\Lambda^2}+ {\cal O}(|S|^4/\Lambda^4)\right)+|\lambda S \bar{q}\pm M\bar{q}|^2+|\lambda S q \pm M q|^2
\end{equation}
and it has a susy minimum at 
\begin{equation} 
S=\mp \frac{M}{\lambda}, \; \; q \bar{q}=\frac{\mu^2}{\lambda},
\end{equation}
and a susy  breaking minimum at 
\begin{equation} \label{MNnS}
S=q=\bar{q}=0. 
\end{equation}
The $S$ gets a mass and is stabilized at the origin due to the non-canonical terms in the K\"ahler potential. Loops interactions with messengers, which do not respect the $U(1)_R$ because of the mass term, generate the following Coleman-Weinberg effective potential for $S$ \cite{Murayama:2007fe}:
\begin{equation}
V_{CW} \simeq \frac{5\mu^4}{(4\pi)^2}\left(\frac{\lambda^3}{M}(S+S^\dagger)-\frac{\lambda^4}{2M^2}(S^2+S^{\dagger\, 2})+...\right)
\end{equation}
The meta-stability of the ($\ref{MNnS}$) susy breaking vacuum can be checked by looking at the mass matrices of the $S$, $q$ and $\bar{q}$:
\begin{equation}
m^2_S \simeq \left(\begin{array}{cl}      
4\frac{\mu^4}{\Lambda^2} &  -\frac{5\mu^4\lambda^4}{(4\pi)^2M^2}  \\     
-\frac{5\mu^4\lambda^4}{(4\pi)^2M^2} & 4\frac{\mu^4}{\Lambda^2}   \\     
\end{array}\right),
\,\,\,\,\,\,\,\,\,\,\,\,\,\,\,\,\,\,\,\,\,\,\,\,\,\,\, m^2_q=
%\begin{math}
\left(\begin{array}{cl}      
|\lambda S \pm M|^2 & -\lambda \mu^2  \\     
-\lambda \mu^2 & |\lambda S \pm M|^2   \\     
\end{array}\right)
\end{equation}
\\ 
at the origin. Radiative corrections due to the coupling of the spurion to messengers can render the mass of $S$ tachyonic unless $M> \lambda^2\Lambda/(4\pi)$. The $q$, $\bar{q}$ directions are stable as long as $M^2>\lambda\mu^2$ is satisfied. Otherwise one of the messengers becomes tachyonic, the susy breaking vacuum disappears and the model becomes effectively the one given by  ($\ref{KitanoW}$)-($\ref{KitanoK}$). The exchange of messengers gives rise to the gaugino masses of the order of \cite{Dine:1994vc}
\begin{equation}
m_{1/2} \simeq \frac{\alpha}{4\pi}\frac{\lambda \mu^2}{M}
\end{equation}
where $\alpha$ represents a generic standard model gauge coupling. The fact that this model lacks a $U(1)_R$ symmetry is the reason why it is claimed to give viable phenomenology while having a metastable susy breaking vacuum at the origin contrary to models that respect $U(1)_R$ \cite{Cheung:2007es}. 

We shall demonstrate  that the inclusion of gravity changes drastically the vacuum structure. The dominant terms in the  scalar potential read
\begin{eqnarray} \nonumber
&V_0\simeq -2\frac{c}{M^2_P}\mu^2 (S+S^\dagger)+4\mu^4 \frac{|S|^2}{\Lambda^2}- \lambda\mu^2(q\bar{q}+q^\dagger\bar{q}^\dagger)-2 \mu^2 \frac{c}{M^2_P} (S+S^\dagger) (|q|^2+|\bar{q}|^2)&  \\ \nonumber 
&+|\lambda S \pm M|^2 (|q|^2+|\bar{q}|^2)+\lambda^2 |q|^2|\bar{q}|^2.&  \nonumber
\end{eqnarray}
It has been taken into account that the dimensionful constant $c/M_P$ must be of the order of the susy breaking scale $\mu^2$ for the vanishing cosmological constant. A few remarks are in order here. Firstly, the susy breaking minimum is shifted from the origin $S=0$ to the non zero value $S\simeq c\Lambda^2/(2M^2_P\mu^2)$. Secondly, since $\Lambda \geq M$,  the $S \sim \Lambda^2/M_P$ can be close to the susy preserving vacuum, although it is easy to arrange $\Lambda^2/M_P \ll M/\lambda$. 
Hence, this model has the interesting feature that the susy breaking minima are closer to the origin compared to the susy preserving ones. This fact raises the question whether the susy breaking minima are thermally preferred. In order to check this we write the finite temperature potential assuming a temperature higher than the messenger scale $M$
\begin{equation} \nonumber
V= V_0+V^T_1=V_0+\frac{T^2}{12}\left[4\frac{\mu^4}{\Lambda^2}+4\mu^2\frac{c}{M^2_P}\frac{S+S^\dagger}{\Lambda^2}+ 3|\lambda S\pm M|^2+3\lambda^2(|q|^2+|\bar{q}|^2)\right]+ {\cal O}(T^4).
\end{equation}
At temperatures $T>\Lambda$ the thermal average field values are $q=\bar{q}=0$ and $S=\mp M/\lambda + {\cal O}(\mu^2c/(M^2_P\lambda^2\Lambda^2))$. As the temperature decreases the mass squared of messengers at this minimum, taking $q=\bar{q}$, is 
\begin{equation}
m^2_q\simeq -2\lambda \mu^2 +2|\lambda S \pm M|^2 +\frac{1}{2}T^2\lambda^2 \simeq -2\lambda \mu^2 +\frac{1}{2}T^2\lambda^2.
\end{equation}
Therefore, the situation is similar to those presented in the previous chapters. If the susy breaking sector and messenger are in thermal equilibrium with the thermal bath then the system at the critical temperature $T_{cr} \simeq 2\mu/ \sqrt{\lambda}$ will evolve to the phenomenologically unacceptable susy preserving vacuum. The situation could change if we assume further couplings of $S$ of the form $\delta W=\lambda' S\phi\bar{\phi}$ where $\phi, \; \bar{\phi}$ are fields uncharged under the SM gauge group \cite{Katz:2009gh}. This could shift the high temperature minimum of $S$ closer to the origin and change the thermal history. However, the  evolution of the system becomes  then highly model dependent. We note that the idea of adding to the superpotential a term $\delta W=\lambda' S\phi\bar{\phi}$ was first presented in the paper by Ellis, Llewellyn Smith and Ross \cite{Ellis:1982vi} where their mezzo-O'Raifeartaigh model was modified in this manner to push the thermal minimum towards the susy breaking metastable vacuum. 

As explained in the previous section, an inflationary phase is expected to displace the fields towards the region  of relatively large vevs. The results of that section can be also applied in the present  case with the important difference that here the susy breaking and susy preserving minima may exchange their roles in the arguments. For instance,
if the vev  of the spurion after inflation gets shifted into the vicinity of the susy preserving minimum, then the system can find itself to be trapped there.

\section{Comparison to ISS}

It is interesting to compare the general situation which we have found in models with  K\"ahler potential stabilization to that known from the widely studied case of the ISS model. ISS showed that metastable susy breaking is possible in a wide class
of remarkably simple models.  One of their main examples is
supersymmetric $SU(N_c)$ QCD with $N_f$ flavours.  If one lies
in the free magnetic range, $N_c<N_f<\frac32 N_c$, then the low energy
theory is strongly coupled, but admits a dual interpretation in terms
of IR-free, magnetic variables.

The tree-level superpotential in the magnetic theory is given by:
\begin{align} \label{ISS}
  W_{\mbox{tree}}=h\mbox{Tr}\bigl(\phi \Phi \tilde{\phi}\bigr) - h \mu^2
  \mbox{Tr}(\Phi)
\end{align}
where $\Phi$ transforms as $N_f \times \overline{N_f}$, $\phi$:
$(\overline{N_f},N)$, $\tilde{\phi}$: $(N_f,\overline{N})$,
$N=N_f-N_c$ is the number of squark colours in the magnetic theory and we
denote the parts of $\Phi$ that  obtain expectation values
as follows: $\Phi=\left(\begin{array}{cc}
    \Phi_1 & 0 \\
    0 & \Phi_0\end{array} \right)$.   The K\"ahler potential is canonical. 

Considering the tree-level superpotential in isolation one finds that
the lowest energy state is a moduli space parameterized by

\begin{equation}
  \Phi=\left( \begin{array}{cc}
      0 & 0 \\
      0 & \Phi_0\end{array} \right), \quad \quad \phi= \left(\begin{array}{c} \phi_0 \\ 0 \end{array} \right), \quad \quad \tilde{\phi}^{T}= \left(\begin{array}{c} \tilde{\phi}_0 \\ 0 \end{array} \right), \quad \quad \phi_0 \tilde{\phi}_0 = \mu^2 \, \mathbb{I}_{N_c \times N_c}. \end{equation}
Supersymmetry is broken by the rank condition. When the one-loop effects
are included the moduli space is lifted and, aside from flat
directions identified with Goldstone bosons, a unique minimum is found
at:
\begin{equation} \Phi=0, \quad \quad\phi_0=\tilde{\phi_0}=\mu \,
  \mathbb{I}_{N_c \times N_c}. \end{equation}

In addition one must include the non-perturbative, R-symmetry
violating contribution:
\begin{align} \label{nnpert}
  W = Nh^{N_f/N}\bigl(\Lambda_m^{-(N_f-3N)}\det(\Phi)\bigr)^{1/N}.
\end{align}
Notice that the exponent of $\Lambda_m$,
$-\left(N_f-3N\right)=-(3N_c-2N_f)$, is always negative in the free
magnetic range.  Hence the coefficient of the determinant grows as the
cut-off shrinks. Since the non-perturbative piece is R-symmetry violating a susy preserving
minimum does exist.  

Lets assume that the ISS hidden sector is thermalized. At high enough temperatures the origin is the minimum of the finite temperature effective potential. In the meson direction far away from the origin a second minimum (which becomes the susy preserving one at $T=0$) forms, but it is always seperated by a barrier from the minimum at the origin. At a temperature $T^{\text{ssb}}_c \sim \mu$ the curvature of the potential at the origin becomes negative in the quark direction but stays positive in the meson direction. A new minimum forms in the quark direction, a phase transition occurs and the fields move to the newly formed minimum. At a temperature $T^{susy}_c \sim (h\mu^2)^{1/2}$ the isolated minimum in the meson direction becomes degenerate with the one at the origin. The potential barrier between them implies that the  transition could be accomplished through quantum tunneling between the two vacua which  is much more strongly suppressed than the classical  transition in the quark direction. As the temperature decreases the minimum in the meson direction becomes the global one and the other minimum, close to the origin in the quark direction, becomes metastable. The minima are always seperated by a potential barrier. The conclusion is that the transition to the non-susy vacuum is thermally favored. Also, it was shown in \cite{Abel:2006cr,Abel:2006my} that even if the fields start in the supersymmetric minimum, e.g. due to non-adiabatic initial conditions, high enough temperatures will thermally drive them to the susy breaking minimum. In particular, if the reheating temperature is $T_{rh}> {\cal O} (1) \, \mu $ the universe ends up in the non-susy vacuum.

The ISS evolution is exactly the opposite of what has been  discussed in the previous chapters, where susy vacua were thermally preferred. Large thermal masses stabilize the fields at the origin of field space. At such small vevs of the fields the non-perturbative piece which creates the supersymmetry preserving vacuum is irrelevant\footnote{The overall power of the components of $\Phi$ in the nonperturbative piece is $N_f /(N_f - N_c)$, which is larger than 2 given the validity range of the model.}. It is the tree level superpotential which determines the behaviour near the origin, enhanced by thermal corrections. The basic difference between the models considered earlier and the  ISS is  due to the multi-field structure of ISS and to the rank condition breaking, which relegates the supersymmetric vacuum from the vicinity of the origin.  As a result, the number of light degrees of freedom is larger near the origin, i.e. near the nonsupersymmetric vacuum. Hence, as the temperature drops, the closest, that  is supersymmetry breaking, minimum is naturally selected in the case of ISS. In O'Raifeartaigh models studied here the situation is different - at high temperatures the corrections which are responsible for the stabilization of the spurion at the supersymmetry breaking minimum are irrelevant, and the supersymmetric minimum gets naturally selected.

However, it should be noted, that the models studied in this work belong to the class known as ordinary gauge mediation with explicit messengers and with K\"ahler potential stabilization, whereas in the original ISS analysis explicit messengers have not been considered. In the literature there are several deformations of the ISS model with ordinary or direct gauge mediation. An example of explicit messenger sector  added to the ISS (\ref{ISS}) is $W_\text{mess}=-\lambda \text{Tr}(\Phi)q\bar{q}+Mq\bar{q}$ \cite{Murayama:2006yf,Abel:2009ze}. In principle one could imagine a large number of messengers (at least of the order of $N\times N_f$ sets of messengers) which become light far from the origin, for instance due to the presence of an explicit mass term. Although the squarks of the ISS sector are light at $\Phi = 0$,  the actual physical masses of messengers  could vanish at $\Phi \sim M/\lambda$.  Then the thermal minimum, preferred by the large number of light states, could be at $\Phi\neq 0$, contrary to the previous conclusions concerning  the pure ISS sector.  Taking also into account that the presence of messengers increases the number of susy preserving vacua in the field space, this could result in thermal selection of a phenomenologically wrong vacuum. However, such a setup  is non-generic.

\section{Comment on moduli, gravity and anomalous $U(1)$}

If one considers stringy models then the moduli of various kinds enter the game (KKLT compactification models with an uplifting ISS or O'Raifairtaigh sector at finite temperature were considered e.g. at \cite{Papineau:2008xf, Anguelova:2007ex}). The are two distinct ways moduli would modify the Lagrangian \cite{Jelinski:2009cp}. 
Firstly, and most interestingly, the linear term in the superpotential may acquire the modular factor:
\begin{equation}
\mu^2 S \rightarrow \mu^2 e^{- T} S,
\end{equation}
where both the Polonyi field $S$ and the modulus $T$ are charged under an anomalous $U(1)$ symmetry. The additional D-term contribution to the scalar potential would appear as well, of the form 
\begin{equation}
V_D = \frac{1}{2} \left ( T + \bar{T} - |S|^2 \right )^2.
\end{equation}
The exact analysis of the stabilization of the fields in such a case is  model dependent, but in general can be achieved. If this is the case, we do not expect a significant change to the postinflationary cosmological evolution of the models we have analyzed. 
The point is that the modulus which tranforms non-linearly will form the longitudinal degree of freedom of the 'anomalous' gauge boson. As such the whole modulus supermultiplet together with the massless vector multiplet will form a massive vector multiplet which should decouple from the low energy dynamics. In other words, we expect that the modulus will become very massive (as massive as the additional $U(1)$ gauge boson) and will not affect the low energy postinflationary dynamics. The other effect of the modulus will be a Planck scale suppressed modification of mass terms, but this effect should be subdominant, of the order of $m_{3/2}$.

\section{Summary and conclusions}

Gauge mediation of supersymmetry breakdown has many attractive features and can be realized in phenomenologically interesting string-motivated models. We have studied the thermal and non-thermal evolution of the coupled susy breaking and messenger sectors. It is demonstrated  that in models with a spurion field stabilized at a low expectation value by quantum and/or gravitational corrections the low energy metastable supersymmetry breaking vaccum appears to be cosmologically disfavoured if the spurion is not initially displaced sufficiently far away from the origin. 
However, if the spurion at the time of reheating is shifted to the region $\mu/\sqrt{\lambda}< |S|<\Lambda$, the system  can settle into the supersymmetry breaking minimum given the reheating temperature $T_\text{rh}<\text{max}\{T_q, T_S\}$, where $T_q=m_q/20$ is the temperature at which the messengers decouple and, $T_S$ is the temperature at which the metastable vacuum appears. In this letter we have shown that thermalized messengers allow  the selection of the metastable vacuum  provided that the messenger sector is sufficiently weakly coupled to the spurion field. 

The temperature $T_S$ is inversely  proportional to the coupling $\lambda$, whereas the tree level mass of messengers $m_q$ depends linearly on $\lambda$. Therefore, for sufficiently small coupling the $T_S$ can be higher than the messenger mass and for a reheating temperature $m_q<T_\text{rh}<T_S$ the thermalized messengers don't rule out the selection of the metastable minimum. Furthermore, even if  the reheating temperature is higher  than $T_S$ the realization of the supersymmetry breaking vacuum can still be achieved with a sufficiently small coupling $\lambda$. In particular, if $\lambda<T_S$ the thermally induced mass of the spurion, $\lambda T$, is small compared to the Hubble expansion rate. The Hubble friction lets the spurion roll only after $T_S$, i.e. only after the metastable vacuum has already formed. Small values of the coulping $\lambda$ are reasonable if the spurion field is a composite operator above the scale $\Lambda$ as is often the case in dynamical supersymmetry breaking scenarios.

We have shown that asymmetric (biased) initial conditions for the spurion, displacing it  far away from the origin can easily be implemented and should be considered natural in the framework of inflationary cosmology. Moreover, we examined the evolution of the fields after inflation and before reheating and we have argued that  even the tracing of the metastable minimum by the spurion could be arranged, minimizing the possibility that the oscillations of $S$ could  overclose the universe. The fact that the reheating temperature can take rather high values without excluding the supersymmetry breaking vacuum selection leaves space for the  leptogenesis  to be realized. 

To present an example, for a theory with a cut-off scale $\Lambda=10^{16}$GeV, coupling $\lambda=10^{-8}$ and assuming that the spurion finds itself in (or sufficiently close to) the metastable susy breaking vacuum before reheating, then, the sysytem of fields will select the metastable vacuum without any constraint on the reheating temperature. Cosmological problems related to the big bang nucleosynthesis resulting from late decays of NLSPs or overclosure of the universe by  the gravitino LSP  can also be avoided in the same manner as in generalized gauge mediation models \cite{Olechowski:2009bd}.

To summarize, we have shown that the requirements of the metastable vacuum selection don't jeopardize the low energy phenomenology. A small coupling between the spurion field, i.e. the supersymmetry breaking sector, and the messengers stabilizes the metastable vacuum at relatively large values $\left\langle S\right\rangle\sim 10^{-(4-6)}M_{P}$ where, however, the contribution of gravity to the mediation of supersymmetry breakdown is subdominant and graviton has a mass in the GeV range, which allows  for the standard postinflationary cosmological evolution.

In addition, it has been found that deforming the models by a supersymmetric mass term for messengers in such a way that the susy breaking minimum and the susy preserving minima are all far away from the origin does not change the features of the thermal evolution. On the other hand, in case of the pure ISS hidden sector, where the perturbative susy breaking minimum is close to the origin, initial displacement enhances the chance that the susy preserving minimum becomes selected during the evolution, especially when there is a relatively large number of messengers that become light away from the origin. 
The basic observations  are expected to hold also in the case of stabilized models with anomalous $U(1)$ groups. 
\\
\\
\section*{Acknowledgments}
\vspace*{.5cm}
\noindent This work was partially supported by the EC 6th FrameworkProgramme MRTN-CT-2006- 035863 and by Polish Ministry for Science and Education under grant N N202 091839. ID and  ZL  thank CERN Theory Division  for hospitality. 

\appendix
\section{Effective potential at finite temperature} \normalsize

The two basic assumptions of Sections 4 and 5 are i) that after inflation all the sectors: the visible one, the messenger sector and the secluded (hidden) sector have been reheated very efficiently by inflaton decays and ii) the fields are not displaced from the symmetric thermal minimum.
As we saw in Section 6, this might not be the case. 

The messengers $q,\bar{q}$ interact directly with the electroweak gauge bosons, which are light, hence their averaged interaction rate with the Standard Model fields is of the order of $\alpha_{SM}^{2} T$, which is larger than the expansion rate for temperatures smaller than $10^{14} - 10^{15}$ GeV. However, mesengers have a mass of the order of $m_q = \lambda \langle S \rangle$, which is typically  large with respect to the Fermi scale. According to the standard lore, see \cite{Mukhanov:2005sc}, for the temperatures smaller than $T_q = m_q/ 20$ the messengers decouple from the expansion and become irrelevant as the source of thermal corrections to the potential for $S$. However, as long as messengers are in equilibrium, they contribute thermal corrections to the effective potential, as they couple directly to the spurion field via the term $\lambda^2 S^2 q^2$. 
Even if the spurion $S$ is not in thermal equilibrium with the heat bath, these corrections are there, as the excited messengers with high energies also interact with $S$. In fact, $S$ interacts with the SM particles as well, with the strength which is suppressed by their coupling to the messengers and by the messenger propagators. Finally, the spurions may communicate with the Standard Model via the additional gauge boson, such as the "anomalous" $U(1)$ gauge boson which appears in string-derived models. Although this additional gauge boson is typically only one or two orders of magnitude lighter than the string scale,
it may be sufficient to bring the hidden sector into equilibrium for a period of time. In any case, we shall assume for now that the equilibrium at least for the messengers holds down to the low temperatures $T_q \sim m_q/20$. 
In Section 4 the thermal evolution of the model with gravitational stabilization of the spurion  field will be examined, to answer the question whether the metastable dynamical susy breaking can be naturally realized in the cooling universe. In Section 5 these considerations will be upgraded to a general situation of K\"ahler potential stabilization of the spurion.  In the rest of this section some basic features of finite temperature field theory, necessery for this paper, are presented.

At zero temperature, the expectation value $\phi_c$ of a scalar field $\phi$ is determined by minimizing the effective potential $V(\phi_c)$. One writes the effective potential in the form
\begin{equation}
V(\phi_c)=V_0(\phi_c)+V_1(\phi_c),
\end{equation}
where $V_0$ is the tree level contribution and $V_1$ the one loop correction.
At finite temperature, see e.g. \cite{Quiros:1999jp} the effective potential  $\bar{V}(\phi_c)$ takes the form 
\begin{equation}
\bar{V}(\phi_c)=\bar{V}_0(\phi_c)+\bar{V}_1(\phi_c)
\end{equation}
where $\bar{V}_0$ and $\bar{V}_1$ are the tree level and one loop terms and the expectation value $\phi_c$ is now a thermal average. It is convenient to seperate the one-loop terms into the temperature independent part $\bar{V}^0_1$ (which is identical in form to $V_1$) and the temperature dependent part $\bar{V}^T_1$
\begin{equation}
\bar{V}_1(\phi_c)=\bar{V}^0_1(\phi_c)+\bar{V}^T_1(\phi_c)
\end{equation}
In general for a theory involving scalar fields $\phi_i$, gauge fields $A^\mu_a$ and Weyl fermions $\psi_r$, with the eigenvalues of the mass-squared matrices  $(M^2_S)_i$, $(M^2_V)_a$ and $(M_F)^2_r$, the temperature-dependent one-loop term in the effective potential $\bar{V}^T_1$ takes the form
\begin{equation} \nonumber
\bar{V}^T_1(\phi_c)=\frac{T^4}{2\pi^2} \int^{\infty}_{0} dy y^2 \left \{\sum_{i}\text{ln}\left[1-\text{exp}\left(-\sqrt{y^2+(M^2_S)_i/T^2}\right)\right] \right .
\end{equation}
\begin{equation} \nonumber
+\sum_{a}\left(3 \text{ln} \left[1-\text{exp}\left(-\sqrt{y^2+(M^2_V)_a/T^2}\right)\right]-\text{ln}(1-e^{-y})\right)
\end{equation}
\begin{equation}
\left . -2\sum_{r}\text{ln}\left[1+\text{exp}\left(-\sqrt{y^2+(M_F)^2_i/T^2}\right) \right]  \right \} .
\end{equation}
There are two limits in which $\bar{V}^T_1$ is particularly simple. First, in the limit where all mass-squared eigenvalues are much larger than $T^2$ 
all terms in $\bar{V}^T_1$ approach zero exponentially and $\bar{V}^T_1$ becomes negligible. Second, in the high temperature limit where $T^2$ is much larger than the mass-squared eigenvalues we can write
\begin{equation} \nonumber
\bar{V}^T_1(\phi_c)\simeq -\frac{\pi^2 T^4}{90} \left(N_B+\frac{7}{8}N_F\right)+\frac{T^2}{24}\left[\sum_{i}(M^2_S)_i+3\sum_{a}(M^2_V)_a+ \sum_{r}(M_F)^2_r\right]
\end{equation} 
\begin{equation} \nonumber
-\frac{T}{12\pi}\left[\sum_{i}(M^3_S)_i+3\sum_{a}(M^3_V)_a\right]+...
\end{equation}
\begin{equation} \nonumber
= -\frac{\pi^2 T^4}{90} \left(N_B+\frac{7}{8}N_F\right)+\frac{T^2}{24}\left[trM^2_S(\phi_c)+3trM^2_V(\phi_c)+ trM^2_F(\phi_c)\right]
\end{equation} 
\begin{equation} \label{V-T}
-\frac{T}{12\pi}\left[tr\{M^2_S(\phi_c)\}^{3/2}+3tr\{M^2_V(\phi_c)\}^{3/2}\right]+... \, .
\end{equation}
We are going to suppress the contribution of the vector fields, since some of them are heavier than the messenger scale $\lambda \langle S \rangle$ and the remaining ones give subdominant contributions in the range of scales that we consider. The term proportional to $T^4$ is just the free energy density for an ideal ultra relativistic gas. $N_B$ and $N_F$ are respectively the number of bosonic and fermionic degrees of freedom. If some fields are heavy and some light at  the temperature $T$, then $N_B$ and $N_F$ should be interpreted as the degrees of freedom of light fields, and the traces over the mass matrices should be evaluated only for light fields, since heavy fields do not contribute.

In the case of a supersymmetric theory with non-canonical K\"ahler potential (and including gravity) the traces of the mass matrices are given by the following formulae \cite{Binetruy:1984yx}
\begin{equation} \label{sca-mass}
TrM^2_s=2  K^{i\bar{j}}\frac{\partial^2 V_0}{\partial\phi^i\partial\ \bar{\phi}^{\bar{j}}}
\end{equation}
and
\begin{equation} \label{ferm-mass}
TrM^2_f=  e^G\left[K^{i\bar{j}}K^{k\bar{l}}(\nabla_iG_k+G_iG_k)(\nabla_{\bar{j}}G_{\bar{l}}+G_{\bar{j}}G_{\bar{l}})-2\right] .
\end{equation}
The term $-2$ in (\ref{ferm-mass}) takes into account the mixed goldstino-gravitino contribution. We adopt the notation $\nabla_i G_j=G_{ij}-\Gamma^k_{ij}G_k$ with the connection 
\begin{equation}
\Gamma^{k}_{ij}=K^{k\bar{l}}\partial_i K_{j\bar{l}}.
\end{equation}
In the text we use simplified notation and write $V(\phi)$ instead of $\bar V^{T} (\phi_c)$ etc.

\end{document}